\documentclass{mathincs}

\pdfoutput=1

\usepackage{amsfonts}
\usepackage{amsmath}
\usepackage{amssymb}
\usepackage{amsthm}
\usepackage[english]{babel}
\usepackage[bookmarks=true,colorlinks=true,linkcolor=blue,citecolor=blue,urlcolor=blue]{hyperref}
\usepackage{booktabs}
\usepackage{calc}
\usepackage{caption}
\usepackage{etoolbox}
\usepackage{fancybox}
\usepackage{fancyhdr}
\usepackage[T1]{fontenc}
\usepackage{float}
\usepackage{flushend}
\usepackage[para]{footmisc}
\usepackage{footnote}
\usepackage{graphicx}
\usepackage{ifthen}
\usepackage{listings}
\usepackage[bracketinterpret,seqinfers,footnotecalculus,abbrseqcontext]{logic}
\usepackage{mathrsfs}
\usepackage{longtable}
\usepackage{subcaption}
\usepackage[setrelation]{math}
\usepackage[pretest]{progreg}
\usepackage[bracketinterpret]{dL}
\usepackage[only,shortrightarrow,boxdot,boxempty]{stmaryrd}
\usepackage{pgfopts}
\usepackage{pgfplots}
\usepackage{prettyref}
\usepackage{subcaption}
\usepackage{tabularx}
\usepackage{tikz}
\usepackage{tikz-uml}
\usepackage{xcolor}
\usepackage{xspace}
\usepackage{xstring}

\usetikzlibrary{arrows}
\usetikzlibrary{calc}
\usetikzlibrary{fit}
\usetikzlibrary{positioning,shadows}
\usetikzlibrary{automata}

\DeclareRobustCommand{\Abbreviation}[2]{#1.\,#2.\xspace}
\newcommand{\eg}{\Abbreviation{e}{g}}
\newcommand{\ie}{\Abbreviation{i}{e}}

\newcommand{\wrt}{w.r.t.\xspace}

\DeclareRobustCommand{\todo}[2][\empty]{%%% \empty: default of the optional argument
  \ifthenelse{\equal{#1}{\empty}}
    {\textsf{\textbf{\color{red}Todo: #2}}}
    {\textsf{\color{red}(#1)\textbf{ Todo: #2}}}
}
\DeclareRobustCommand{\comment}[2][\empty]{%%% \empty: default of the optional argument
  \ifthenelse{\equal{#1}{\empty}}
    {\textsf{\textbf{\color{green}Comment: #2}}}
    {\textsf{\color{green}(#1)\textbf{ Comment: #2}}}
}

\providecommand{\norm}[1]{\lVert#1\rVert}

\DeclareRobustCommand{\Sphinx}[0]{Sphinx\xspace}
\providecommand{\KeY}{Ke\kern-0.1emY\xspace}

\providecommand{\UmlSt}[1]{\guillemotleft #1\guillemotright}

\newcommand{\rref}[2][]{\prettyref{#2}}
\newrefformat{sec}{Section\,\ref{#1}}
\newrefformat{def}{Def.\,\ref{#1}}
\newrefformat{thm}{Theorem\,\ref{#1}}
\newrefformat{prop}{Proposition\,\ref{#1}}
\newrefformat{lem}{Lemma\,\ref{#1}}
\newrefformat{ex}{Example\,\ref{#1}}
\newrefformat{tab}{Table\,\ref{#1}}
\newrefformat{fig}{Fig.\,\ref{#1}}
\newrefformat{model}{Model\,\ref{#1}}
\newrefformat{proof}{Proof\,\ref{#1}}
\newrefformat{rewrite}{(\ref{#1})}

\DeclareRobustCommand{\Qte}[1]{\flqq #1\frqq\xspace}

\tikzset{
    state/.style={
    		circle,
           draw=black, thick
           },
	plant/.style={
           rectangle,
           rounded corners,
           draw=black, thick,
           minimum height=2em,
           inner sep=2pt,
           text centered,
           },
}

\tikzstyle{accepting}=[accepting by arrow]

\tikzumlset{fill class=white, fill note=white, fill state=white, fill package=gray!20, font=\footnotesize}
\renewcommand{\tikzumlstatedotext}{}

\floatstyle{ruled}
\newfloat{model}{tbp}{model}
\floatname{model}{Model}

\newtheorem{thm}{Theorem}%[section]

\newlength{\linferenceRulehskipamount}

\newcommand{\rightrule}{r}
\newcommand{\leftrule}{l}
\newfloat{proof}{tbp}{proof}
\floatname{proof}{Proof}

\newcounter{tabline}
\newcommand{\tabrn}[1][1]{\refstepcounter{tabline}\thetabline}

\graphicspath{{./}{fig/}}

\begin{document}

% BEGIN IEEE
\title[Collaborative Verification-Driven Engineering of Hybrid Systems]{Collaborative Verification-Driven Engineering of\\ Hybrid Systems}

\author{Stefan Mitsch}
\address{Computer Science Department, Carnegie Mellon University\\
and Department of Cooperative Information Systems, Johannes Kepler University\\
5000 Forbes Ave,\\
Pittsburgh, PA 15213, USA}
\email{smitsch@cs.cmu.edu}

\author{Grant Olney Passmore}
\address{LFCS, Edinburgh and Clare Hall, Cambridge\\ 10 Crichton Street,\\ Edinburgh, UK}
\email{grant.passmore@cl.cam.ac.uk}

\author{Andr{\'e} Platzer}
\address{Computer Science Department, Carnegie Mellon University\\
5000 Forbes Ave,\\
Pittsburgh, PA 15213, USA}
\email{aplatzer@cs.cmu.edu}

%\jourtitle{Math.Comput.Sci.}
%\DOInr{10.1007/s11786-014-0176-y}
%\volume{}
%\onlinedate{27 April 2014}
%\copyrightinfo{2014}{Springer Basel}
%\copyrightyear{2014}
%\copyrightholder{Springer Basel}

\begin{abstract}
  Hybrid systems with both discrete and continuous dynamics are an
  important model for real-world cyber-physical systems.  
  The key challenge
  is to ensure their correct functioning \wrt safety
  requirements.  
  Promising techniques to ensure safety seem to be model-driven engineering to develop hybrid systems in a well-defined and traceable manner, and formal verification to prove their correctness. Their combination forms the vision of verification-driven engineering. 
  Often, hybrid systems are rather complex in that they require expertise from many domains (\eg, robotics, control systems, computer science, software engineering, and mechanical engineering).
  Moreover, despite the remarkable progress in automating
  formal verification of hybrid systems, the construction of proofs of
  complex systems often requires nontrivial human guidance, since
  hybrid systems verification tools solve undecidable problems.
  It is, thus, not uncommon for development and verification teams to consist of many
  players with diverse expertise.  This paper introduces a
  verification-driven engineering toolset that extends our previous
  work on hybrid and arithmetic verification with tools for (i)
  graphical (UML) and textual modeling of hybrid systems, (ii) exchanging and comparing models and
  proofs, and (iii) managing verification tasks.  This toolset makes
  it easier to tackle large-scale verification tasks.
\end{abstract}



%\thanks{This material is based upon work supported by the National Science
%Foundation under NSF CAREER Award CNS-1054246, NSF EXPEDITION
%CNS-0926181, and under Grant Nos.  CNS-1035800 and CNS-0931985, by
%DARPA under agreement number FA8750-12-2-0291, and by the US
%Department of Transportation's University Transportation Center's TSET
%grant, award\# DTRT12GUTC11. Passmore was also supported by the UK's
%EPSRC [grants numbers EP/I011005/1 and EP/I010335/1].
%This work was also supported by the Austrian Federal Ministry of Transport, Innovation and 
%Technology (BMVIT) under grant FFG FIT-IT 829598, FFG BRIDGE 838526, and FFG Basisprogramm 
%838181.
%The research leading to these results has received funding from the People Programme 
%(Marie Curie Actions) of the European Union's Seventh Framework Programme (FP7/2007-2013) 
%under REA grant agreement n$^\circ$ PIOF-GA-2012-328378.}

\pagestyle{headings} 

\lhead{Math.Comput.Sci. (2014)\\DOI 10.1007/s11786-014-0176-y} 
\rhead{}
\lfoot{}
\rfoot{}

\maketitle

\setcounter{footnote}{0}

\thispagestyle{fancy}

\noindent Received: 1 November 2013 / Revised: 15 February 2014 / Accepted: 3 March 2014\\
\copyright Springer Basel 2014\\
\noindent A journal version of this article appeared with Springer \cite{DBLP:journals/mics/MitschPP14} and is available at link.springer.com:\\ \noindent\footnotesize{\url{http://link.springer.com/article/10.1007/s11786-014-0176-y}}

% main text
\section{Introduction}
\label{sec:intro}

Computers that control physical processes form so-called \emph{cyber-physical systems} (CPS), which are pervasively embedded into our lives today. 
For example, cars equipped with adaptive cruise control form a typical CPS~\cite{DBLP:conf/fm/LoosPN11} that is responsible for controlling acceleration on the basis of distance sensors.
Further prominent examples can be found in many safety-critical areas, such as in factory automation~\cite{DBLP:conf/ki/NiemuellerERKFJL13}, medical equipment~\cite{DBLP:journals/pieee/LeeSCHJKKMPRV12}, power plants and grid~\cite{DBLP:journals/pieee/SridharHG12}, automotive~\cite{DBLP:conf/hybrid/DeshpandeGV96,DBLP:conf/samos/GoswamiSMLCVA12}, aviation~\cite{Tomlin1998}, and railway industries~\cite{DBLP:conf/icfem/PlatzerQ09}. %\cite{DBLP:conf/icfem/PlatzerQ09,DBLP:conf/fm/LoosPN11,DBLP:conf/fm/PlatzerC09}.
From an engineering viewpoint, a CPS can be described as a \emph{hybrid system} in terms of discrete control decisions (the cyber-part, \eg, setting the acceleration of a car) and differential equations modeling the entailed physical continuous dynamics (the physical part, \eg, motion)~\cite{DBLP:journals/jar/Platzer08}. 
More advanced models of CPS include aspects of distributed hybrid systems or stochasticity~\cite{DBLP:conf/lics/Platzer12a}, but are not addressed here.

The key challenge in engineering hybrid systems is the question of how to ensure their correct functioning in order to avoid incorrect control decisions \wrt safety requirements (\eg, a car with adaptive cruise control will never collide with a car driving ahead).
Promising techniques to ensure safety seem to be \emph{model-driven engineering} (MDE)~\cite{DBLP:conf/ifm/Kent02,DBLP:journals/computer/Schmidt06} to incrementally develop systems in a well-defined and traceable manner and formal verification to mathematically prove their correctness. 
Together, these techniques form the vision of \emph{verification-driven engineering} (VDE)~\cite{Kordon2008}.

Often, CPS are complex systems and require expertise in many different domains, for instance, in robotics, control systems, computer science, software engineering, and mechanical engineering.
Despite the remarkable progress in automating formal verification of hybrid systems, many interesting and complex verification problems still remain that are hard to solve in practice with a single tool by a single person.
It is, thus, not uncommon for serious hybrid systems development and
verification teams to consist of many players, some with expertise in
robotics, control theory and dynamical systems, some in software
engineering, some in mathematical logic, some in real algebraic geometry, and so on.
Hence, modeling languages that convey a model to a broad and possibly heterogeneous audience together with integrated tools in a toolchain and well-established project management techniques to coordinate team members are crucial to achieve effective collaborative large-scale verification of hybrid systems (cf., \eg,~\cite{DBLP:journals/tse/CraigenGR95,Woodcock:2009:FMP:1592434.1592436}).
For example, the graphical nature of UML seems to be suitable to convey information even to UML novices and promote communication between team members (\eg, as observed in a large industry study for safety-critical process control~\cite{DBLP:journals/ese/AndaHGT06}).
Moreover, it can even increase comprehensibility and accessibility of formal notations~\cite{Razali:2007:CUF:1353673.1353680}.

Collaboration on CPS verification is important for yet another reason: Because hybrid systems are undecidable~\cite{DBLP:journals/tcs/AlurCHHHNOSY95}, hybrid systems verification tools work over an undecidable theory, and so verifying complicated systems within them often requires significant human guidance. 
This need for human guidance is true even for \emph{decidable} theories utilized within hybrid systems verification~\cite{Collins2005}, such as the first-order theory of non-linear real arithmetic (also called the theory of \emph{real closed fields} or RCF), a crucial component of real-world verification efforts. 
Though decidable, RCF is fundamentally infeasible (it is worst-case doubly exponential~\cite{Davenport1988}), which poses a problem for the automated verification of hybrid systems.  
Much expertise is needed to discharge arithmetical verification conditions in a reasonable amount of time and space, expertise requiring the use of deep results in real algebraic geometry. 
Successful examples of team-based large-scale verification of non-hybrid systems include the operating system kernel seL4~\cite{Klein2009} in Isabelle/HOL and the Flyspeck project~\cite{DBLP:journals/dcg/HalesHMNOZ10}, and show that, indeed, collaboration is key for proving large systems.
Similar effects are expected in CPS verification.

This paper introduces the VDE toolset \Sphinx comprising modeling and verification tools for hybrid systems (including a backend deployment for project management and collaboration support).
The toolset applies proof decomposition in-the-large across multiple verification tools, basing on the completeness of differential dynamic logic (\dL~\cite{DBLP:journals/jar/Platzer08,DBLP:conf/lics/Platzer12b}), which is a real-valued first-order dynamic logic for hybrid programs, a program notation for hybrid systems.
\Sphinx extends our previous work on the deductive
verification tool \KeYmaera~\cite{DBLP:conf/cade/PlatzerQ08} and on
the nonlinear real arithmetic verification tools RAHD
\cite{PassmorePhD} and MetiTarski~\cite{DBLP:conf/aisc/PassmorePM12} with tools for (i) graphical (UML) modeling, model transformation, and textual modeling of hybrid systems, (ii) exchanging and comparing models and proofs, and (iii) exchanging knowledge and tasks through a project management backend.

\noindent\textbf{Structure of the paper.}
In the next section, we give an overview of related work.
In~\rref{sec:vde} we introduce our architecture of a verification-driven engineering toolset, and describe implementation and features of its components.
\rref{sec:application} introduces an autonomous robotic ground vehicle as application example.
Finally, in~\rref{sec:conclusion} we conclude the paper with an outlook on real-world application of the toolset and possible directions for future work.

\section{Related Work}
\label{sec:relatedwork}

Model-driven engineering in a collaborative manner has been successfully applied in the embedded systems community.
Efforts, for instance, include transforming between different UML models and SysML~\cite{Hause:2008:IMA:1396379.1396417}, modeling in SysML and transforming these models to the simulation tool Orchestra~\cite{Bajaj2012}, integration of modeling and simulation in Cosmic/Cadena ~\cite{DBLP:journals/scp/GokhaleBKBEDTPS08}, or modeling of reactive systems and integration of various verification tools in Syspect~\cite{FLO+2011}.

Recent surveys on verification methods for hybrid systems~\cite{DBLP:conf/emsoft/Alur11}, modeling and analysis of hybrid systems~\cite{Schutter2009}, and modeling of cyber-physical systems~\cite{Derler2012}, reveal that indeed many tools are available for modeling and analyzing hybrid systems, but in a rather isolated manner.
Supporting collaboration on formal verification by distributing tasks among members of a verification team in a model-driven engineering approach has not yet been the focus.
Although current verification tools for hybrid systems (\eg, PHAVer~\cite{DBLP:conf/hybrid/Frehse05}, SpaceEx~\cite{FrehseLGDCRLRGDM11}), as well as those for arithmetic (\eg, Z3~\cite{DBLP:conf/tacas/MouraB08}) are accompanied by modeling editors of varying sophistication, they are not yet particularly well prepared for collaboration either.
Developments in collaborative verification of source code by multiple complementary static code checkers~\cite{Wuestholz2012}, modular model-checking (\eg,~\cite{Kupferman:1997:MMC:646738.702091}), and extreme verification~\cite{DBLP:conf/birthday/HenzingerJMS03}, however, indicate that this is indeed an interesting field.
Most notably, usage of online collaboration tools in the Polymath project has led to an elementary proof of a special case of the density Hales-Jewett theorem~\cite{Gowers2009}.

The Unified Modeling Language (UML~\cite{Hitz2005}) is a standardized language for object-oriented modeling. 
But without extension it is not well suited for modeling hybrid systems~\cite{DBLP:journals/sttt/BerkenkotterBHP06}.
Therefore, the profiling mechanism of UML was used to extend the standardized UML languages SysML~\cite{Hause:2008:IMA:1396379.1396417} for modeling hardware and software components of complex systems and MARTE~\cite{DBLP:conf/simutools/MalletS08} for modeling real-time and embedded systems.
These and other extensions~\cite{DBLP:journals/sttt/BerkenkotterBHP06,DBLP:journals/fcsc/LiuLHMD13,DBLP:conf/birthday/SchaferW10} increase the support for hybrid modeling in UML. 
However, those profiles augment the UML Statechart formalism, since their languages base on hybrid automata as underlying principle.
We, instead, use hybrid programs and therefore extend UML Activity Diagrams since they are a more natural way of modeling.
Examples for integrating formal notations with informal ones can be found outside the hybrid systems community, for instance, UML-B~\cite{DBLP:journals/tosem/SnookB06}, TRIO~\cite{DBLP:conf/sigsoft/LavazzaQV01}, or VeriAgent \cite{DBLP:journals/entcs/MotaCGOFK04}.

In summary, we address model-driven engineering and formal verification as follows.
\begin{itemize}
\item Unlike~\cite{Hause:2008:IMA:1396379.1396417,Bajaj2012,DBLP:journals/scp/GokhaleBKBEDTPS08}, who focus on exchanging models, we also facilitate collaboration on formal verification.
\item Unlike~\cite{DBLP:conf/hybrid/Frehse05,FrehseLGDCRLRGDM11,DBLP:conf/tacas/MouraB08}, who focus on one aspect of verification, we provide modeling and collaboration tools that should make it easier for domain experts to work in verification teams and exchange models and verification results between different tools.
\item Unlike~\cite{DBLP:conf/hybrid/Frehse05,DBLP:conf/tacas/MouraB08}, who focus on verification tools, we also work on modeling support and collaboration.
\item Unlike~\cite{DBLP:journals/sttt/BerkenkotterBHP06,DBLP:journals/fcsc/LiuLHMD13,DBLP:conf/birthday/SchaferW10}, who define a hybrid automaton semantics for UML Statecharts, we define a hybrid program semantics for UML Activity Diagrams.
\item
Unlike~\cite{DBLP:conf/sigsoft/LavazzaQV01,DBLP:journals/entcs/MotaCGOFK04,DBLP:journals/tosem/SnookB06}, who combine formal models with semi-formal modeling in UML for discrete systems, we define a UML profile for hybrid systems modeling.
\end{itemize}

\section{The VDE Toolset \Sphinx}
\label{sec:vde}

\begin{figure}[t!b]
%\centering
\includegraphics[width=\textwidth]{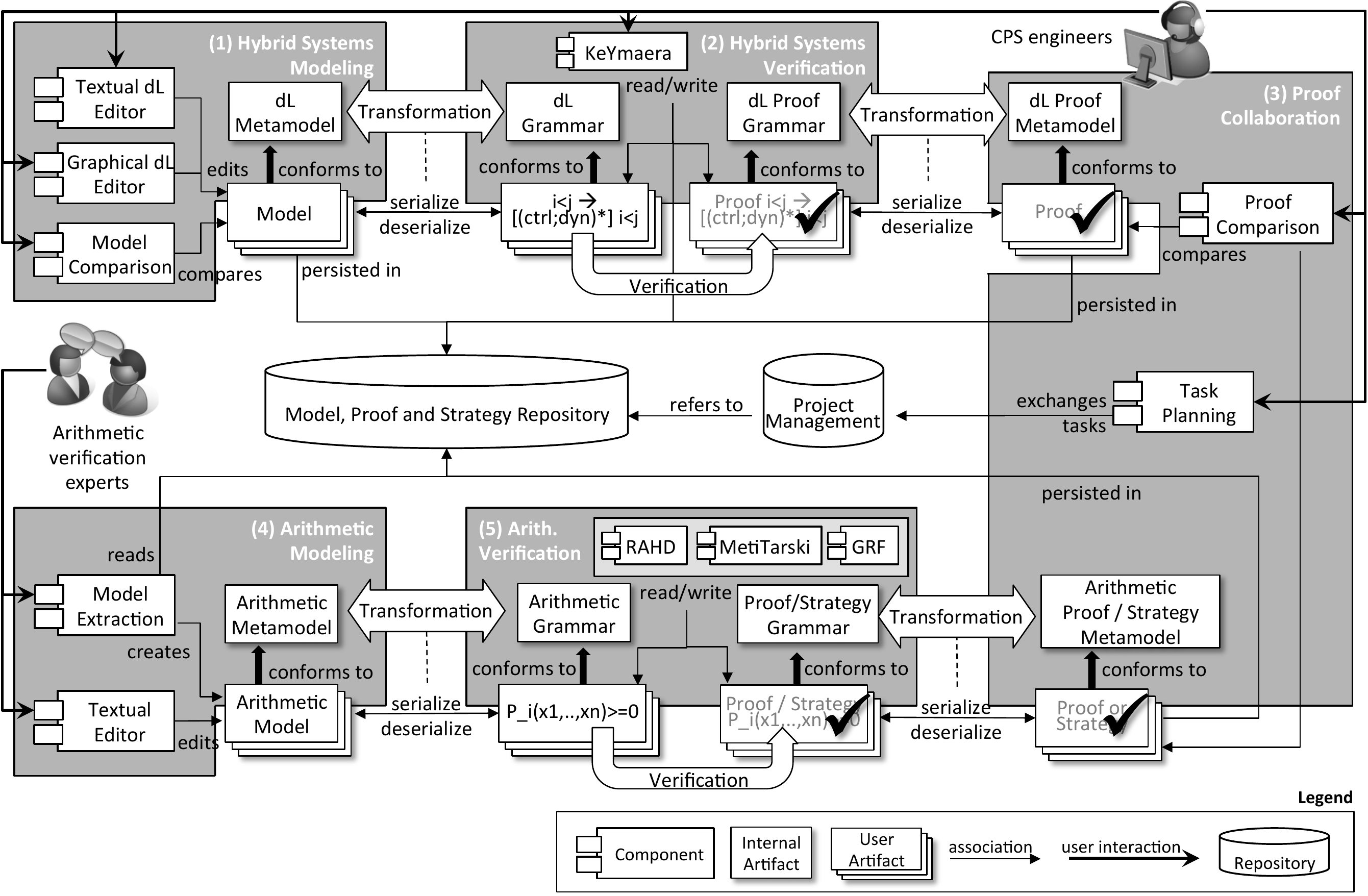}%{architecture2.pdf}
\caption{Overview of components in the verification-driven engineering toolset. 
The toolset provides components for (1) hybrid systems modeling, (2) hybrid systems verification, (3) proof collaboration, (4) arithmetic modeling, and it provides or uses off-the-shelf components for (5) arithmetic verification.
It uses further off-the-shelf components as repositories in the backend.}
\label{fig:architecture}
\end{figure}

Verification teams often comprise experts with diverse heterogeneous background who are accustomed to different modeling and verification tools with heterogeneous notations.
In order to integrate different modeling and verification tools, the verification-driven engineering toolset \Sphinx\footnote{\url{http://www.cs.cmu.edu/~smitsch/sphinx.html}} proposed in this paper follows a model-driven architecture: metamodels for different modeling and proof languages form the basis for manipulating, persisting, and transforming models.
The idea is to provide general-purpose graphical and textual modeling languages for hybrid systems, while at the same time keeping the \Sphinx platform open for additional languages.
That way, we can still develop and integrate domain-specific languages (DSL), which are specifically tailored to the terminology used in a particular domain and with a semantics defined by transformation to hybrid programs~\cite{DBLP:journals/jar/Platzer08,DBLP:journals/logcom/Platzer10,Platzer10,DBLP:conf/lics/Platzer12b}.
Such DSLs would enable domain experts to express models that are suitable for verification purposes in their familiar terminology.

The notion of a model here denotes an instance of a metamodel, \ie, it comprises hybrid system models, proofs, and strategies.
Following the definition of the Object Management Group (OMG\footnote{\url{http://www.omg.org}}), a metamodel defines a language to formulate models: one example for a metamodel is the grammar of differential dynamic logic (\dL~\cite{DBLP:journals/jar/Platzer08,DBLP:journals/logcom/Platzer10,Platzer10,DBLP:conf/lics/Platzer12b},), which, among others, defines language elements for non-deterministic choice, sequential composition, assignment, repetition, and differential equations.
An example for a model is given in~\rref{sec:vde-watertank}: it describes a simple water tank as a set of formulas, differential equations, and other \dL language elements.
The model conforms to the grammar of \dL, and thus is an instance of the \dL metamodel.
\rref{fig:architecture} gives an overview of the toolset architecture:
the \dL metamodel, \dL proof metamodel, arithmetic metamodel, and arithmetic proof metamodel each represent an interface between tools and to the backend.

\begin{description}
\item[\dL metamodel] The hybrid modeling components (textual and graphical editors for \dL, as well as model comparison) manipulate models that conform to the \dL metamodel.
The \dL models are serialized to and deserialized from their textual form that can be read by \KeYmaera.
\item[Hybrid Program UML] The hybrid program UML profile extends UML with hybrid system concepts that can be translated to \dL models.
\item[\dL proof metamodel] The proof comparison component reads proofs that conform to the \dL proof metamodel.
These proofs may either be closed ones (completed proofs, nothing else to be done) or partial proofs (to be continued).
Again, proofs in \Sphinx are serialized and deserialized from the textual form as generated by \KeYmaera.
\item[Arithmetic metamodel] Arithmetic editors (not yet implemented) manipulate arithmetic models. 
Again transformations are performed between models expressed in terms of the arithmetic metamodel and the corresponding textual input (\eg, SMT-LIB syntax~\cite{Barrett2012}) as needed by arithmetic tools, such as RAHD~\cite{PassmorePhD}, MetiTarski~\cite{DBLP:journals/jar/AkbarpourP10}, Z3~\cite{DBLP:conf/tacas/MouraB08}, GRF~\cite{ppz:grf:2013}, or MathSat~\cite{DBLP:conf/tacas/CimattiGSS13}.
\item[Arithmetic proof metamodel] Finally, the proof comparison component reads arithmetic proofs expressed in terms of the arithmetic proof metamodel, which is serialized to and deserialized from the textual format of the arithmetic tool.
\end{description}

\subsection{A Hybrid Water Tank Example}
\label{sec:vde-watertank}

\begin{figure}[htb]
\centering
\begin{subfigure}[b]{\textwidth}
\centering
\begin{footnotesize}
\begin{tikzpicture}
\umlclass[type=System]{World}
	{+ $\varepsilon$ : $\mathbb{R}$ \{readOnly\}\\
	 + c : $\mathbb{R}$	
	}
	{}
\umlclass[type=Agent,x=6]{Watertank}
	{+ x : $\mathbb{R}$	\\
	 + f : $\mathbb{R}$\\
	 + M : $\mathbb{R}$ \{readOnly\}\\
	}
	{}

\umlunicompo[geometry=-|-]{World}{Watertank}

\umlnote[x=9,y=-0.5,width=15ex]{Watertank}{\vspace{-2ex}\begin{center}\UmlSt{Invariant}\\\{dL\} $M \geq 0$\end{center}}
\umlnote[x=3,y=-0.7,width=14ex]{World}{\vspace{-2ex}\begin{center}\UmlSt{Invariant}\\\{dL\} $\varepsilon > 0$\end{center}}
\end{tikzpicture}
\end{footnotesize}
\caption{The structure of the water tank model: the current water level $x$ must not exceed the maximum level $M$ when the flow $f$ is chosen once every $\varepsilon$ time units, which will be triggered by the clock $c$.}
\label{fig:graphical-watertankstructure}
\end{subfigure}
\begin{subfigure}[b]{\textwidth}
\begin{footnotesize}
\begin{tikzpicture}
\umlstateinitial[name=initial,y=0.5]
\umlstatedecision[name=loopbegin,x=1,y=0.5]
%\begin{umlstate}[name=ctrl,x=3]{ctrl}

%\end{umlstate}
\umlbasicstate[name=ctrl,x=3,type=X,do=$\humod{f}{*}$]{\UmlSt{AssignAny} ctrl}
\umlbasicstate[name=dyn,x=9.3,width=28ex,do=\protect{$\humod{c}{0};~ \left(\D{x} = f ~\&~ c \leq \varepsilon \land x \geq 0\right)$}]{\UmlSt{Dynamics} dyn}
\umlstatedecision[name=loopend,x=12,y=0.5]
%\umlstatedecision[name=loopskip,x=12.7]
%\umlstatefinal[name=final,x=13.5]
\umlstatefinal[name=final,x=13,y=0.5]

\umltrans{initial}{loopbegin}
\umltrans{loopbegin}{ctrl}
\umltrans[name=thetest]{ctrl}{dyn}
\umltrans{dyn}{loopend}
\umltrans{loopend}{final}
%\umltrans{loopend}{loopskip}
%\umltrans{loopskip}{final}
\umlrelation[style=<->,geometry=|-|, arm1=-2cm,name=loop,stereo=NondetRepetition,pos stereo=1.5]{loopend}{loopbegin}
%\umlVHVtrans[arm1=-1.5cm,name=loop,stereo=NondetRepetition,pos stereo=1.5]{loopend}{loopbegin}

\umlnote[x=1.5,y=-2.5,width=30ex]{initial}{\vspace{-2ex}\begin{center}\UmlSt{Initial}\\\{dL\} $0 \leq x \leq M\land \varepsilon > 0$\end{center}}
\umlnote[x=12,y=-2.5,width=20ex]{final}{\vspace{-2ex}\begin{center}\UmlSt{Safety}\\\{dL\} $0 \leq x \leq M$\end{center}}
\umlnote[x=8,y=-2.5,width=25ex]{loop-3}{\vspace{-2ex}\begin{center}\UmlSt{Invariant}\\\{dL\} $0 \leq x \leq M$\end{center}}
\umlnote[x=5.7,y=-0.5,width=17ex]{thetest-1}{\vspace{-2ex}\begin{center}\UmlSt{Test}\\\{dL\} $f \leq \frac{M - x}{\varepsilon}$\end{center}}
\end{tikzpicture}
\end{footnotesize}
\caption{The behavior of the water tank model: the controller $\textit{ctrl}$ chooses a new nondeterministic value for the flow $f$, such that it satisfies the subsequent test before it is passed to the continuous dynamics $\textit{dyn}$. 
Controller and continuous dynamics are repeated nondeterministically many times (which means it can be skipped entirely, cf.~\rref{tab:hybridprograms} on page \pageref{tab:hybridprograms} and the hybrid program UML semantics on page \pageref{sec:hpumlhpsemantics}).}
\label{fig:graphical-watertankbehavior}
\end{subfigure}
\caption{Example of a hybrid system: a water tank}
\label{fig:graphical-watertank}
\end{figure}
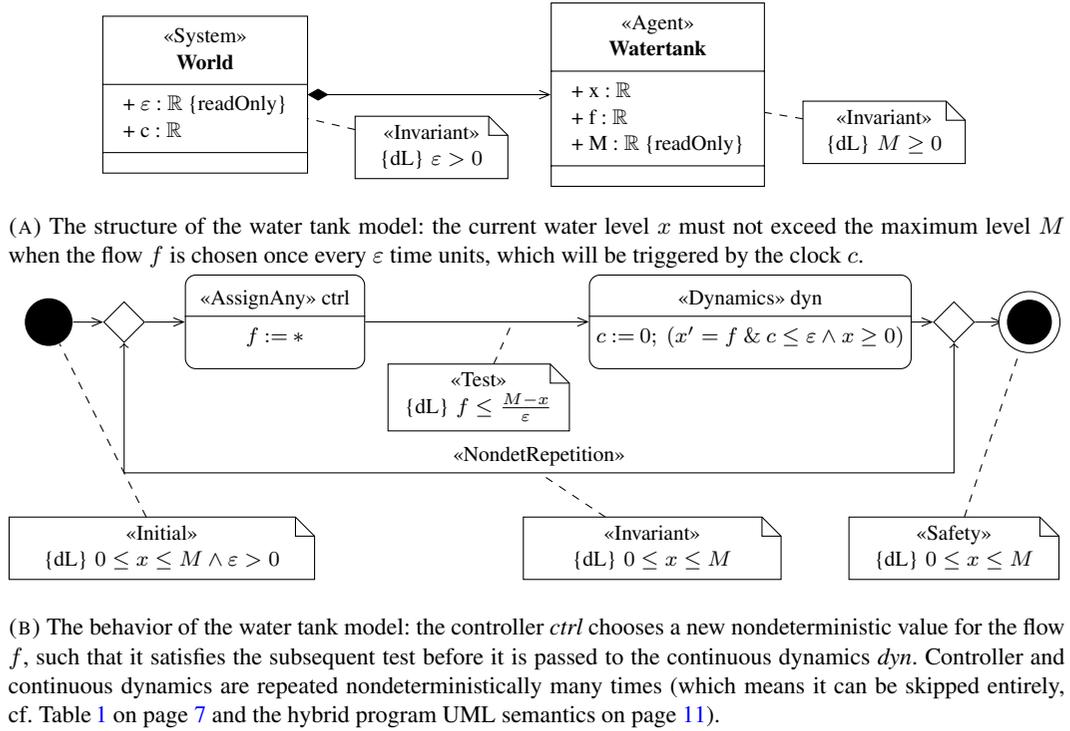

We illustrate the notion of hybrid systems and our hybrid programs and hybrid program UML profile by means of the classical water tank example: a water tank should not overflow when the flow in or out of the water tank is chosen once every time interval.
The hybrid program UML model is shown in~\rref{fig:graphical-watertank}.
We will use the hybrid program UML syntax informally here and later introduce it in detail in~\rref{sec:hybridprogramuml}.

The system introduces a global clock $c$ and a bound on the loop execution time $\varepsilon$, which must be strictly positive as indicated by the invariant $\varepsilon > 0$ attached to the class $\textit{World}$.
The system further consists of one agent, the $\textit{WaterTank}$, which is characterized by the current water level $x$, the current flow $f$ and the maximum level $M$.
The maximum level is constant ($\textit{readOnly}$) and non-negative, as defined in the attached invariant $M \geq 0$.

The behavior of the system is a single loop with two actions: the $\textit{ctrl}$ action chooses a new nondeterministic flow that will not exceed the water tank's maximum capacity (cf. the test $\tfrac{M-x}{\varepsilon}$). 
The subsequent continuous evolution $\textit{dyn}$ resets the clock $c$ and evolves the water level in the tank according to the chosen flow along the differential-algebraic equation $\D{x}=v ~\&~ c \leq \varepsilon \land x \geq 0$ (the constraints ensure that the clock will not exceed a certain limit $c \leq \varepsilon$ and the water level will always be non-negative $x \geq 0$).

The specification about the water tank is annotated as constraints on the initial and the final node: when we start the water tank model in a state where the current level is within the limits of the water tank, then all runs of the model should keep the water level within the limits.

\subsection{Development and Verification Process Overview}

Let us exemplify the \Sphinx toolset with a virtual walk through a collaborative verification scenario.
We begin with modeling a hybrid system as in the water tank example above using the graphical and textual \dL editors.
The resulting model, which conforms to the \dL metamodel, is transformed on-the-fly during editing to a textual input file, and loaded into \KeYmaera.
In \KeYmaera, we apply various strategies for proving safety of our hybrid system model, but may get stuck at some difficult arithmetic problem.
We mark the corresponding node in the partial proof and save it in \KeYmaera's textual output format.
The proof collaboration tool transforms the partial proof text file into a model of the partial proof. 
We persist the hybrid model and the model of the partial proof in the model and proof repository.
Then we create a request for arithmetic verification (ticket) in the project management repository using the task planning component.
The assignee of the ticket accesses the linked partial proof, and extracts an arithmetic verification model from the marked proof node.
Then a transformation runtime creates the textual input for one of the arithmetic verification tools.
In this tool, a proof for the ticket can be created, along with a proof strategy that documents the proof.
Such a proof strategy is vital for replaying the proof later, and for detecting whether or not the arithmetic proof still applies when the initial model has changed.
Both, proof and proof strategy, are imported into the proof collaboration tool and persisted to the corresponding repository.
The ticket is closed, together with the node on the original proof (if the arithmetic proof is complete; otherwise, the progress made is reported back).
We fetch the new proof model version from the repository and inspect it using the proof comparison component.
Then we transform the proof model into its textual form, load \KeYmaera and continue proving our hybrid system from where we left off, but now with one goal closed.
In case the corresponding arithmetic prover is connected to \KeYmaera, we could even load the proof strategy from the strategy repository and repeat it locally to reduce proof effort on other subgoals.

\subsection{\KeYmaera: Hybrid System Verification}

\KeYmaera\footnote{\url{http://symbolaris.com/info/KeYmaera.html}}~\cite{DBLP:conf/cade/PlatzerQ08} is a verification tool for hybrid systems that combines deductive, real algebraic, and computer algebraic prover technologies.
It is an automated and interactive theorem prover for a natural specification and verification logic for hybrid systems.
\KeYmaera supports differential dynamic logic (\dL)~\cite{DBLP:journals/jar/Platzer08,DBLP:journals/logcom/Platzer10,Platzer10,DBLP:conf/lics/Platzer12b}, which is a real-valued first-order dynamic logic for hybrid programs, a program notation for hybrid systems. 
\KeYmaera supports hybrid systems with nonlinear discrete jumps, nonlinear differential equations, differential-algebraic equations, differential inequalities, and systems with nondeterministic discrete or continuous input.

For automation, \KeYmaera implements a number of automatic proof strategies that decompose hybrid systems symbolically and prove the full system by proving properties of its parts~\cite{Platzer10}.
This compositional verification principle helps scaling up verification, because \KeYmaera verifies a big system by verifying properties of subsystems.
Strong theoretical properties, including relative completeness results, have been shown about differential dynamic logic~\cite{DBLP:journals/jar/Platzer08,DBLP:conf/lics/Platzer12b}.

\KeYmaera implements fixedpoint procedures~\cite{DBLP:journals/fmsd/PlatzerC09} that try to compute invariants of hybrid systems and differential invariants of their continuous dynamics, but may fail in practice.
By completeness~\cite{DBLP:journals/jar/Platzer08,DBLP:conf/lics/Platzer12a,DBLP:conf/lics/Platzer12b}, this is the only part where \KeYmaera's automation can fail in theory.
In practice, however, also the decidable parts of dealing with arithmetic may become infeasible at some point, so that interaction with other tools or collaborative verification via \Sphinx is crucial.

At the same time, it is an interesting challenge to scale to solve larger systems, which is possible according to completeness but highly nontrivial. 
For systems that are still out of reach for current automation techniques, the fact that completeness proofs are compositional can be exploited by interactively splitting parts of the hybrid systems proof off and investigating them separately within \Sphinx.
If, for instance, a proof node in arithmetic turns out to be infeasible within \KeYmaera, this node could be verified using a different tool connected to \Sphinx.

\KeYmaera has been used successfully for verifying case studies from train control~\cite{DBLP:conf/icfem/PlatzerQ09}, car control~\cite{DBLP:conf/itsc/LoosP11,DBLP:conf/fm/LoosPN11,DBLP:conf/iccps/MitschLP12}, air traffic management~\cite{DBLP:conf/hybrid/LoosRP13,DBLP:conf/fm/PlatzerC09}, robotic obstacle avoidance~\cite{DBLP:conf/rss/MitschGP13}, and robotic surgery~\cite{DBLP:conf/hybrid/KouskoulasRPK13}.
These verification results illustrate how some systems can be verified automatically while others need more substantial user guidance.
The \KeYmaera approach is described in detail in a book~\cite{Platzer10}.
\KeYmaera is linked to \Sphinx by implementing extensions to the Eclipse launch configuration.
These extensions hook into the context menu of Eclipse (models in \dL and \dL proof files in our case) and, on selection, launch \KeYmaera as an external program.
In the same fashion, further verification tools can be connected to \Sphinx.

In order to guide domain experts in modeling discrete and continuous dynamics of hybrid systems, the case studies, further examples, and their proofs are included in the \KeYmaera distribution.
When applying proof strategies manually by selection from the context menu in the interactive theorem prover, \KeYmaera shows only the applicable ones sorted by expected utility.
Preliminary collaboration features include marking and renaming of proof nodes, as well as extraction of proof branches as new subproblems.
These collaboration features are used for interaction with the arithmetic verification tools and the collaboration backend described below.

\subsection{Real Arithmetic Verification}

Proofs about hybrid systems often require significant reasoning about
multivariate polynomial inequalities, \ie, reasoning within the
\emph{theory of real closed fields} (RCF). Though
RCF is decidable, it is fundamentally infeasible
(hyper-exponential in the number of variables). It is not uncommon for
hybrid system models to have tens or even hundreds of real variables,
and RCF reasoning is commonly the bottleneck for nontrivial
verifications. Automatic RCF methods simply do not scale, and
manual human expertise is often needed to discharge a proof's
arithmetical subproblems.

RCF infeasibility is not just a problem for hybrid systems
verification. Real polynomial constraints are pervasive throughout the
sciences, and this has motivated a tremendous amount of work on the
development of feasible proof techniques for various special classes
of polynomial systems. In the context of hybrid systems verification,
we wish to take advantage of these new techniques as soon as possible.

Given this fundamental infeasibility, how might one go about deciding
large RCF conjectures? One approach is to develop a battery
of efficient proof techniques for different practically useful
fragments of the theory. For example, if an $\exists$~RCF
formula can be equisatisfiably transformed into an
$\wedge\vee$-combination of strict inequalities, then one can
eliminate the need to consider any irrational real algebraic solutions
when deciding the formula. Tools such as RAHD~\cite{PassmorePhD}, Z3~\cite{DBLP:conf/tacas/MouraB08} and MetiTarski~\cite{DBLP:journals/jar/AkbarpourP10}
exemplify this heterogeneous approach to RCF, and moreover
allow users to define \emph{proof strategies} consisting of heuristic
combinations of various specialized proof methods. When faced with a
difficult new problem, one works to develop a proof strategy which can
solve not only the problem at hand but also other problems sharing
similar structure. Such strategies, though usually constructed by
domain experts, can then be shared and utilized as automated
techniques by the community at large.

As verification tools like KeYmaera progress, they accumulate a large
database of RCF facts which pertain to the system being analyzed.
As subsequent RCF subproblems are generated, they are tested for
validity modulo this database of background facts.
In practice, often only a small subset of the background RCF facts are
needed to decide the generated subproblems.
The difficulty lies in how the most relevant facts should be selected.
The geometric relevance filtering (GRF) method~\cite{ppz:grf:2013} is
an RCF decision method combining high-dimensional sampling techniques
and incremental cell decomposition methods adapted from cylindrical
algebraic decomposition (CAD) to use geometric information to select
relevant background facts.
GRF supports human experts in deciding which arithmetic subproblems to keep and which ones to discharge.

\subsection{Modeling and Proof Collaboration}
\label{sec:sphinx_syntax}

In order to interconnect the variety of specialized verification procedures introduced above, \Sphinx follows a model-driven engineering approach: it introduces metamodels for the included modeling and proof languages.
These metamodels provide a clean basis for model creation, model comparison, and model transformation between the formats of different tools.
This approach is feasible, since in principle many of those procedures operate over the theory RCF, or at least share a large portion of symbols and their semantics.
One could even imagine that very same approach for exchanging proofs between different proof procedures, since proofs in RCF, in theory, can all be expressed in the same formal system.
Currently, proofs in \Sphinx are exchanged merely for the sake of being repeated in the original tool (although \KeYmaera already utilizes many such tools and hence is able to repeat a wide variety of proofs).

\begin{figure}[t!b]
%\centering
\includegraphics[width=\textwidth]{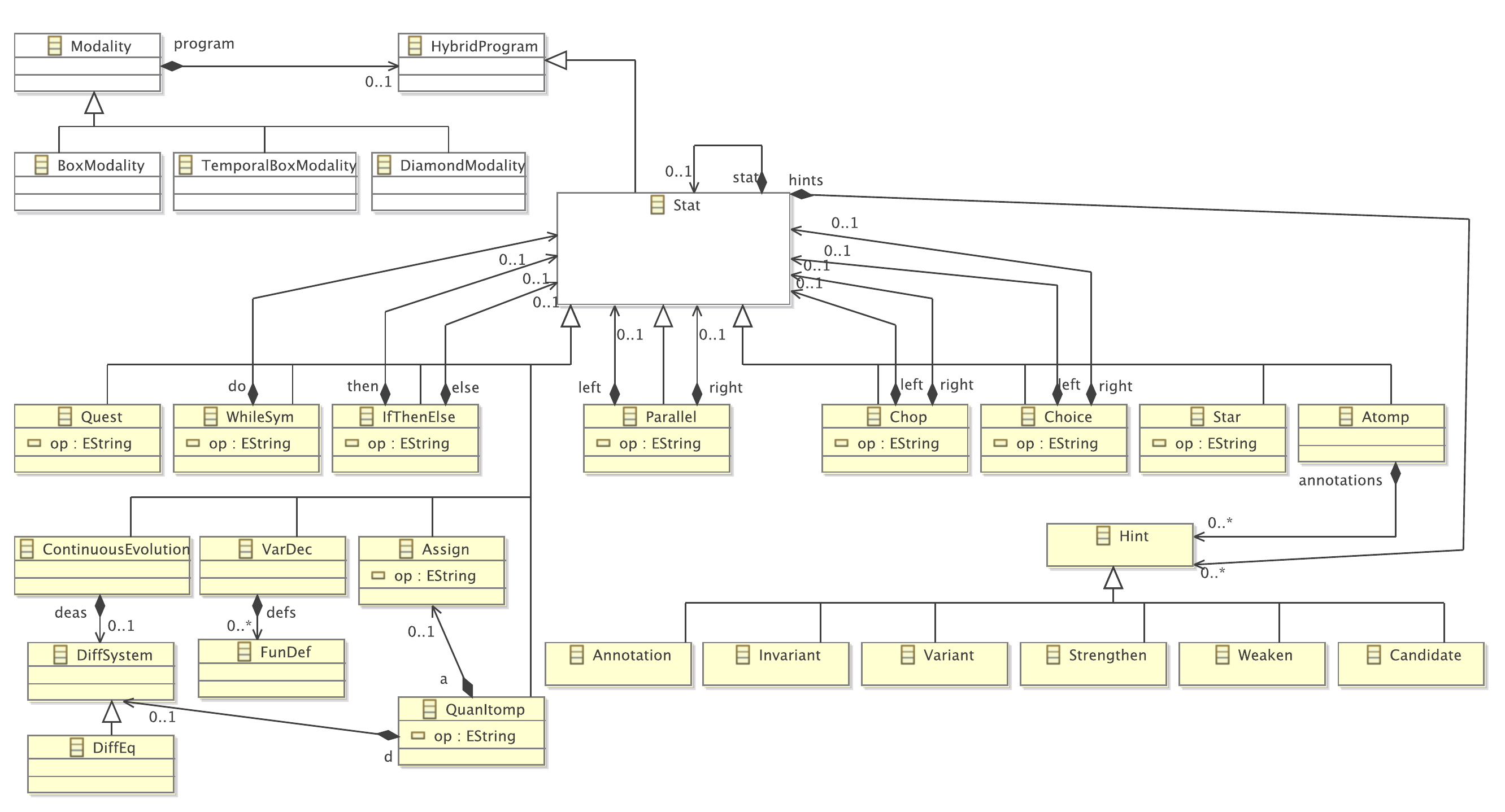}%{DifferentialDynamicLogic.pdf}
\caption{The \dL metamodel extracted from the input grammar of \KeYmaera}
\label{fig:dlmetamodel}
\end{figure}

In the case of textual languages, \Sphinx uses the Eclipse Xtext\footnote{\url{www.eclipse.org/Xtext}} framework to obtain metamodels directly from the language grammars 
(cf.~\rref{fig:dlmetamodel}, obtained from the \dL grammar~\cite{DBLP:journals/jar/Platzer08}), together with other software artifacts, such as a parser, a model serializer, and a textual editor with syntax highlighting, code completion, and cross referencing. 
These metamodels are the basis for creating models in \dL, as well as for defining transformations between \dL and other modeling languages.
The models in \dL make use of mathematical terms, and are embedded in \KeY files since \KeYmaera uses the \KeY~\cite{Mostowski2007} format for loading models and saving proofs.
In the following sections, we introduce \dL in more detail and describe the support for creating \dL models and working on proofs in \Sphinx.

\subsubsection{Differential Dynamic Logic}
\label{sec:dl}

\begin{table}[b!]
  \newcommand{\foform}{F\xspace}
%  \centering
  \caption{Statements of hybrid programs}
  \begin{footnotesize}
  \begin{tabular}{l@{~}|@{~}l@{~}|l@{~}}
    \multicolumn{1}{c@{~}|@{~}}{Statement} & Metamodel element & \multicolumn{1}{c}{Effect}
    \\
    \hline 
    $\alpha;~\beta$ & Chop & sequential composition, performs~HP $\alpha$ and then HP $\beta$ afterwards\\
    $\alpha~\cup~\beta$ & Choice & nondeterministic choice, follows either HP~$\alpha$ or HP~$\beta$\\
    $\alpha^*$ & Star & nondeterministic repetition, repeats HP~$\alpha$ $n\geq 0$ times \\
    $\humod{x}{\theta}$ & Assign (term) & discrete assignment of the value of term $\theta$ to variable $x$ (jump)\\
    $\humod{x}{*}$ & Assign (wild card term) & nondeterministic assignment of an arbitrary real number to $x$\\
    $\bigl(\D{x_1}=\theta_1,\dots,$ & ContinuousEvolution & continuous evolution of $x_i$ along differential equation system\\ 
    \qquad{$\D{x_n}=\theta_n ~\&~ \foform\bigr)$} & DiffSystem & $\D{x_i} = \theta_i$, restricted to maximum domain or invariant region~$\foform$\\
    $?\foform$ & Quest & check if formula $\foform$ holds at current state, abort otherwise\\
%    $\text{if}(\foform) \text{ then } \alpha$ & perform $\alpha$ if $\foform$ holds, do nothing otherwise\\
    $\text{if}(\foform) \text{ then } \alpha \text{ else } \beta$ & IfThenElse & perform HP $\alpha$ if $\foform$ holds, perform HP $\beta$ otherwise\\
    $\text{while}(\foform) \text{ do } \alpha \text{ end}$ & WhileSym & perform HP $\alpha$ as long as $\foform$ holds\\
    $\dbox{\alpha}{\phi}$ & BoxModality & \dL formula $\phi$ must hold after all executions of HP $\alpha$\\
    $\ddiamond{\alpha}{\phi}$ & DiamondModality & \dL formula $\phi$ must hold after at least one execution of HP $\alpha$\\
  \end{tabular}
  \end{footnotesize}
  \label{tab:hybridprograms}
\end{table}

For specifying and verifying correctness statements about hybrid systems, we use \emph{differential dynamic logic} \dL~\cite{DBLP:journals/jar/Platzer08,Platzer10,DBLP:conf/lics/Platzer12b}, which supports \emph{hybrid programs} as a program notation for hybrid systems.
The syntax of hybrid programs is summarized together with an informal semantics in~\rref{tab:hybridprograms}; the metamodel introduced in~\rref{fig:dlmetamodel} reflects this syntax.
The sequential composition \Qte{$\alpha;~\beta$} expresses that $\beta$ starts after $\alpha$ finishes (\eg, first let a car choose its acceleration, then drive with that acceleration). 
The non-deterministic choice \Qte{$\alpha~\cup~\beta$} follows either $\alpha$ or $\beta$ (\eg, let a car decide nondeterministically between accelerating and braking). 
The non-deterministic repetition operator \Qte{$\alpha^*$} repeats $\alpha$ zero or more times (\eg, let a car choose a new acceleration arbitrarily often).
Discrete assignment \Qte{$x:=\theta$} instantaneously assigns the value of the term $\theta$ to the variable $x$ (\eg, let a car choose a particular acceleration), while \Qte{$x:=*$} assigns an arbitrary value to $x$ (\eg, let a car choose any acceleration).
\Qte{$\D{x}=\theta ~\&~ F$} describes a continuous evolution of $x$ within the evolution domain $F$ (\eg, let the velocity of a car change according to its acceleration, but always be greater than zero).
The test \Qte{$?F$} checks that a particular condition expressed by $F$ holds, and aborts if it does not (\eg, test whether or not the distance to a car ahead is large enough).
A typical pattern that involves assignment and tests, and which will be used subsequently, is to limit the assignment of arbitrary values to known bounds (\eg, limit an arbitrarily chosen acceleration to the physical limits of a car, as in $x:=*; ?x\geq 0$).
The deterministic choice \Qte{$\text{if}(F) \text{ then } \alpha \text{ else } \beta$} executes $\alpha$ if $F$ holds, and $\beta$ otherwise (\eg, let a car accelerate only when it is safe; brake otherwise).
Finally, \Qte{$\text{while}(F) \text{ do } \alpha \text{ end}$} is a deterministic repetition that repeats $\alpha$ as long as $F$ holds.

To specify the desired correctness properties of hybrid programs, differential dynamic logic (\dL) provides modal operators~$\dbox{\alpha}{}$ and~$\ddiamond{\alpha}{}$, one for each hybrid program~$\alpha$. 
When $\phi$ is a \dL formula (\eg, a simple arithmetic constraint) describing a state and $\alpha$ is a hybrid program, then the \dL formula \m{\dbox{\alpha}{\phi}} states that all states reachable by $\alpha$ satisfy $\phi$.
Dually, \dL formula~\m{\ddiamond{\alpha}{\phi}} expresses that there is a state reachable by the hybrid program~$\alpha$ that satisfies \dL formula ~$\phi$.
The set of \dL~formulas is generated by the following EBNF grammar
(where~\m{{\sim}\in\{<,\leq,=,\geq,>\}} and~$\theta_1,\theta_2$ are arithmetic expressions in~\m{+,-,\cdot,/} over the reals):
\begin{center}
\(
\phi ::= \theta_1 \sim \theta_2 \mid \neg \phi \mid \phi \wedge \psi \mid \phi \lor \psi \mid \phi \limply \psi \mid \phi \leftrightarrow \psi \mid \forall x \phi \mid \exists x \phi \mid \dbox{\alpha}{\phi} \mid \ddiamond{\alpha} \phi \enspace .
\)
\end{center}
%Thus, besides comparisons (\m{<,\leq,=,\geq,>}), \dL\ allows one to express negations ($\neg \phi$), conjunctions ($\phi \wedge \psi$), universal ($\forall x \phi$) and existential quantification ($\exists x \phi$), as well as the already mentioned state reachability expressions ($\dbox{\alpha}{\phi}$, $\ddiamond{\alpha} \phi$).

\subsubsection{Creating Models}

\begin{figure}[tb]
\includegraphics[width=\columnwidth]{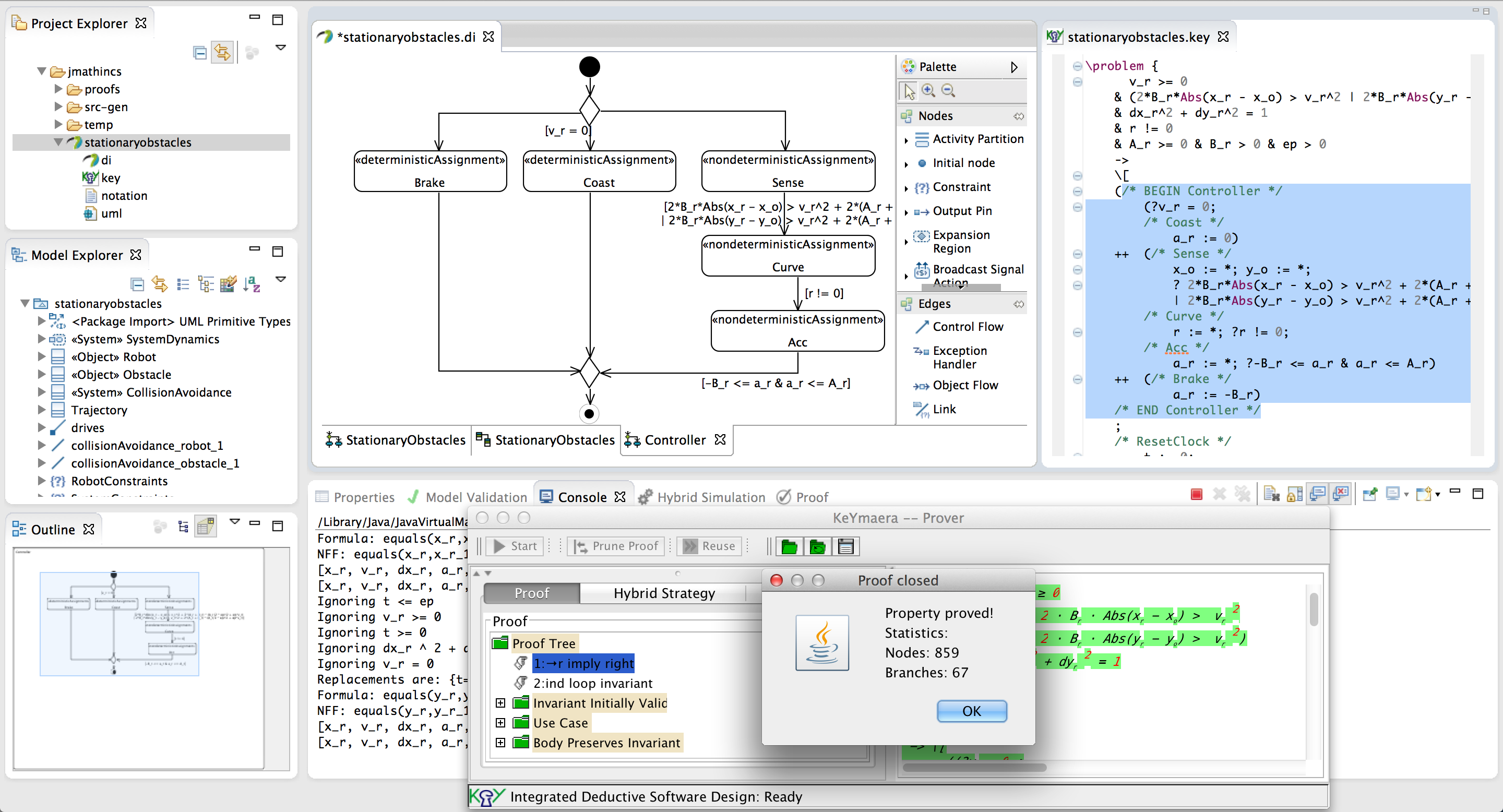}%{screenshot.png}
\caption{Screenshot of the textual and graphical modeling editors and a proof in \KeYmaera (details on the textual and graphical syntax are described in~\rref{sec:sphinx_syntax}).
The left-most three panels, from top to bottom, show the project explorer of Eclipse (access to models and proofs), a hierarchical tree view of the graphical model, and a miniature outline of the graphical model.
The graphical editor is displayed in the center of the tool with the textual editor to the right.
The text editor selection highlights the controller part that is displayed in the graphical editor.
The bottom-most panel shows the \KeYmaera theorem prover with its console output in the background.} 
\label{fig:screenshot}
\end{figure}

\Sphinx currently includes \dL as generic modeling language to create models of hybrid and cyber-physical systems.
The concrete textual \dL editor is created from the \dL metamodel and shown in~\rref{fig:screenshot}, which also illustrates the graphical editor based on UML and the \KeYmaera prover attached through the console.

In order to facilitate the creation of textual models in \dL, \Sphinx includes templates of common model artifacts (\eg, ODEs of linear and circular motion).
These templates, when instantiated, allow in-place editing and automated renaming of the template constituents.
As usual in the Eclipse platform, such templates can be easily extended and shared between team members.

Since generic modeling languages, such as \dL for hybrid systems, tend to incur a steep learning curve, the \Sphinx platform can be extended with dedicated domain-specific languages (DSL).
Such DSLs should be designed to meet the vocabulary of a particular group of domain experts. 
They can be included into \Sphinx in a fashion similar to the generic modeling language \dL, \ie, in the form of Eclipse plugins that provide the DSL metamodel and the modeling editor.

%\paragraph{Validation of \dL Models}
%\todo{Static code analysis}
%\todo{Acknowledgments: Ralph Mayr for implementation}

In the next section we describe the Hybrid Program UML profile, an extension to UML for graphical modeling of hybrid programs that should make it easier to convey the main features of a hybrid system to a broader audience.

\subsubsection{The Hybrid Program UML Profile}
\label{sec:hybridprogramuml}

The Hybrid Program UML profile follows a fundamental principle of UML in that it separates modeling the structure of a hybrid system from modeling its behavior.
Currently, \Sphinx supports class diagrams for modeling the structure of a hybrid system, since hybrid programs do not yet support modules.
Future work includes the addition of composite structure diagrams as used in~\cite{DBLP:journals/sttt/BerkenkotterBHP06,DBLP:conf/birthday/SchaferW10}, and the introduction of proof rules that exploit the additional structural information during the verification process.
For modeling behavior, we use activity diagrams instead of the UML statecharts used in existing hybrid system UML profiles~\cite{DBLP:journals/sttt/BerkenkotterBHP06,DBLP:journals/fcsc/LiuLHMD13,DBLP:conf/birthday/SchaferW10}, since activity diagrams model control flow more akin to hybrid programs.
UML statecharts are a language to model system behavior as graphs where the vertices are the states of the system and the edges represent transitions between states.
Thus, with some extension UML statecharts are suitable to represent hybrid automata (cf.~\cite{DBLP:journals/sttt/BerkenkotterBHP06,DBLP:journals/fcsc/LiuLHMD13,DBLP:conf/birthday/SchaferW10}).
UML activity diagrams, in contrast, are a language to model system behavior as graphs where the vertices are actions or decisions and the edges represent control flow.
The notion of control flow between actions (statements) makes activity diagrams more suitable to model (computational) processes~\cite{Hitz2005}, such as our hybrid programs.

We use model transformation to define the semantics of the Hybrid Program UML profile relative to \dL.
Besides defining the semantics of the Hybrid UML profile, model transformations can be implemented as model transformation specifications (\eg, using the Atlas transformation language ATL~\cite{DBLP:journals/scp/JouaultABK08}) and executed to transform models back and forth between the Hybrid Program UML profile and their hybrid program counterparts.
Since hybrid automata can be encoded in hybrid programs~\cite[Appendix C]{Platzer10},\footnote{
The transformation from hybrid automata into hybrid programs follows the same principle as implementing a finite automaton in a programming language.
The converse transformation from hybrid programs into hybrid automata is based on the transition structure induced by the semantics of hybrid programs~\cite{Platzer10,DBLP:conf/lics/Platzer12a}.
} we define both, a hybrid program semantics and a hybrid automaton semantics, for the Hybrid Program UML profile.
Note, however, that hybrid automata, when encoded in \dL, are often less natural to express and also less efficient to verify than well-structured hybrid programs, because they lack program structure that could be exploited during the proof and require additional variables to identify the states of the automaton.

We use \emph{UML profiles} as extension mechanism to provide hybrid system modeling concepts that are not yet present in standard UML.
Profiles are the standard way to extend UML with domain-specific modeling concepts~\cite{Hitz2005}.
A UML profile is defined by specifying \emph{stereotypes} and \emph{constraints}.
A stereotype is applicable to a particular element of the UML (\eg, a classifier) and adds additional modeling capabilities to the original UML element.
For example, standard UML actions have a body that we can use to capture an atomic hybrid program, such as deterministic assignment or differential equations.
When we want to describe additional information, such as differential invariant constraints or evolution domain constraints of a differential-algebraic equation, 
we can introduce a stereotype \emph{Dynamics} for UML actions that adds the necessary modeling abilities to actions.
Constraints can be added to profiles in order to restrict properties to only admissible values, derive property values from other properties, or otherwise check the consistency of a UML model.
UML provides the Object Constraint Language (OCL,~\cite{OMG2012}) for defining such constraints.
In the following paragraphs we describe our profiles for modeling the structure and the behavior of a hybrid system, which was already informally used in~\rref{sec:vde-watertank}.

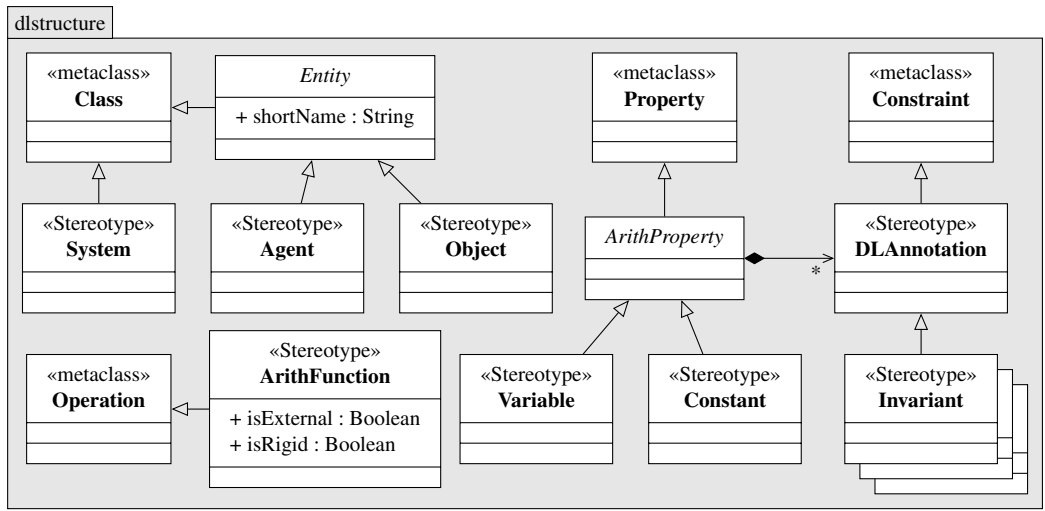
\begin{figure}[htb]
\centering
\begin{footnotesize}
\begin{tikzpicture}
\begin{umlpackage}{dlstructure}
\umlemptyclass[type=metaclass,x=-1.5]{Class}{}{}
\umlemptyclass[type=Stereotype,x=-1.5,y=-2]{System}{}{}
\umlabstract[x=1.5]{Entity}{+ shortName : String}{}
\umlclass[type=Stereotype,x=3.5,y=-2]{Object}{}{}
\umlclass[type=Stereotype,x=1,y=-2]{Agent}{}{}
\umlinherit[geometry=--]{System}{Class}
\umlinherit[geometry=--]{Entity}{Class}
\umlinherit[geometry=--]{Agent}{Entity}
\umlinherit[geometry=--]{Object}{Entity}

\umlemptyclass[type=metaclass,x=6]{Property}{}{}
\umlabstract[x=6,y=-2]{ArithProperty}{}{}
\umlemptyclass[type=Stereotype,x=4.3,y=-4]{Variable}{}{}
\umlemptyclass[type=Stereotype,x=6.8,y=-4]{Constant}{}{}
\umlinherit[geometry=--]{ArithProperty}{Property}
\umlinherit[geometry=--]{Variable}{ArithProperty}
\umlinherit[geometry=--]{Constant}{ArithProperty}

\umlemptyclass[type=metaclass,x=9.4]{Constraint}{}{}
\umlemptyclass[type=Stereotype,x=9.4,y=-2]{DLAnnotation}{}{}
\umlemptyclass[type=Stereotype,x=9.8,y=-4.4]{Generalize}{}{}
\umlemptyclass[type=Stereotype,x=9.6,y=-4.2]{Variant}{}{}
\umlemptyclass[type=Stereotype,x=9.4,y=-4]{Invariant}{}{}
\umlinherit[geometry=--]{DLAnnotation}{Constraint}
\umlinherit[geometry=--]{Invariant}{DLAnnotation}

\umlemptyclass[type=metaclass,x=-1.5,y=-4]{Operation}{}{}
\umlclass[type=Stereotype,x=1.5,y=-4]{ArithFunction}
	{+ isExternal : Boolean\\
	 + isRigid : Boolean
	}{}
\umlinherit[geometry=--]{ArithFunction}{Operation}
	
\umlunicompo[mult=*]{ArithProperty}{DLAnnotation}
\end{umlpackage}

%\umlnote[x=2.5,y=-6, width=3cm]{Class}{Test note}
\end{tikzpicture}
\end{footnotesize}
\caption{The Hybrid Program UML profile: structure of hybrid programs}
\label{fig:profile-dlstatic}
\end{figure}

\paragraph{System Structure}
Hybrid programs in \dL use variables and functions over the reals as modeling primitives.
Further structuring mechanisms, such as classes, are not yet supported in \dL.
In order to still capture and communicate the intended structure of a hybrid program, we provide stereotypes for UML class diagrams\footnote{In principle, a single class would already be a valid structure for a hybrid program. It is, however, useful to split the system into multiple separate classes corresponding to different entities in the system.}.
\rref{fig:profile-dlstatic} shows the stereotypes currently available in the Hybrid UML profile for modeling the structure of a hybrid system.

A hybrid system usually consists of multiple \emph{entities}~\cite{Niles2001}, which are either \emph{objects} that may evolve through manipulation but not by themselves, or \emph{agents} whose evolution is driven by the decisions of a controller (\eg, the robot agent that has to avoid obstacles).

Entities are usually characterized by some \emph{properties} (\eg, the robot's position)~\cite{Baumgartner2014,Kokar2009}.
These properties can be discerned into \emph{constant} properties (cf. \emph{Constant}: their value can be read but not written, \eg, the position of a stationary obstacle), whereas others can change and are therefore called \emph{fluent}~\cite{Reiter2001} or \emph{variable} (cf. \emph{Variable}: their value can both be read and written, \eg, the position of the robot). 
These stereotypes can be equivalently modeled using standard UML notation \emph{readOnly} for properties.
Additional constraints may apply to properties (\eg, a minimum positive braking force $B > 0$, or bounds on the acceleration $-B \leq a \leq A$).
We can model these constraints in the structure of the system using the stereotype \emph{Invariant}, if the constraints have to be satisfied throughout model execution; otherwise, they are part of the system behavior.
We use \dL to formalize those constraints, since OCL does not support arbitrary arithmetic expressions.
Some properties in a hybrid program are shared among all entities (\eg, time).
A class marked with the stereotype \emph{System} can capture such shared knowledge.

We allow further decomposition of agents into multiple classes.
These classes are linked to their respective agent via the \emph{association} concept of UML.
If we want to emphasize that an instance of some class is owned by at most one agent at a time, we use \emph{composition} (\eg, a robot's internal control variables could be factored into a dedicated control state, which no other robot has access to).
If we want to share instances of a class between multiple agents, we use a standard \emph{association} instead.
No further annotation with stereotypes is necessary.

\paragraph{System Behavior}

\begin{figure}[htb]
\centering
\begin{footnotesize}
\begin{tikzpicture}
\begin{umlpackage}{dlbehavior}
\umlemptyclass[type=metaclass,x=1.5]{OpaqueAction}{}{}
\umlabstract[x=-1.5,y=-2]{Assignment}{+ variable : Property}{}
\umlemptyclass[type=Stereotype,x=1.5,y=-2]{Dynamics}{}{}
\umlclass[type=Stereotype,x=-1.8,y=-4]{AssignTerm}{+ term : String}{}
\umlemptyclass[type=Stereotype,x=0.8,y=-4]{AssignAny}{}{}
\umlinherit[geometry=--]{Assignment}{OpaqueAction}
\umlinherit[geometry=--]{Dynamics}{OpaqueAction}
\umlinherit[geometry=--]{AssignTerm}{Assignment}
\umlinherit[geometry=--]{AssignAny}{Assignment}

\umlemptyclass[type=metaclass,x=5]{Constraint}{}{}
\umlemptyclass[type=Stereotype,x=5,y=-2]{DLConstraint}{}{}
\umlemptyclass[type=Stereotype,x=3.5,y=-4]{DiffInvariant}{}{}
\umlemptyclass[type=Stereotype,x=6.3,y=-4]{Test}{}{}
\umlemptyclass[type=Stereotype,x=9,y=-4]{InductiveInvariant}{}{}

\umlemptyclass[type=Stereotype,x=-1.8,y=-6]{Initial}
\umlemptyclass[type=Stereotype,x=0.8,y=-6]{Safety}
\umlemptyclass[type=Stereotype,x=3.5,y=-6]{Liveness}
\umlemptyclass[type=Stereotype,x=6,y=-6]{DiffVariant}
\umlemptyclass[type=Stereotype,x=9,y=-6]{Convergence}
\umlinherit[geometry=--]{DLConstraint}{Constraint}
\umlinherit[geometry=--]{DiffInvariant}{DLConstraint}
\umlinherit[geometry=--]{Test}{DLConstraint}
\umlinherit[geometry=--]{Invariant}{DLConstraint}
\umlinherit[geometry=|-|,arm1=1cm]{Initial}{DLConstraint}
\umlinherit[geometry=|-|,arm1=1cm]{Safety}{DLConstraint}
\umlinherit[geometry=|-|,arm1=1cm]{Liveness}{DLConstraint}
\umlinherit[geometry=|-|,arm1=1cm]{DiffVariant}{DLConstraint}
\umlinherit[geometry=|-|,arm1=1cm]{Convergence}{DLConstraint}

\umlemptyclass[type=metaclass,x=9]{ControlFlow}{}{}
\umlemptyclass[type=Stereotype,x=9,y=-2]{NondetRepetition}{}{}
\umlinherit[geometry=--]{NondetRepetition}{ControlFlow}{}

\umlunicompo[mult=0..1]{Dynamics}{DiffInvariant}
\umlunicompo[mult=0..1]{NondetRepetition}{Invariant}
\umlunicompo[geometry=-|-,anchors=east and east,mult=0..1,pos=2.5,arm1=0.5cm]{NondetRepetition}{Convergence}

\umlnote[x=-1.6,width=15ex]{Assignment}{\{OCL\} not self.isReadOnly}
\end{umlpackage}
\end{tikzpicture}
\end{footnotesize}
\caption{The Hybrid Program UML profile: behavior of hybrid programs}
\label{fig:profile-dldynamics}
\end{figure}
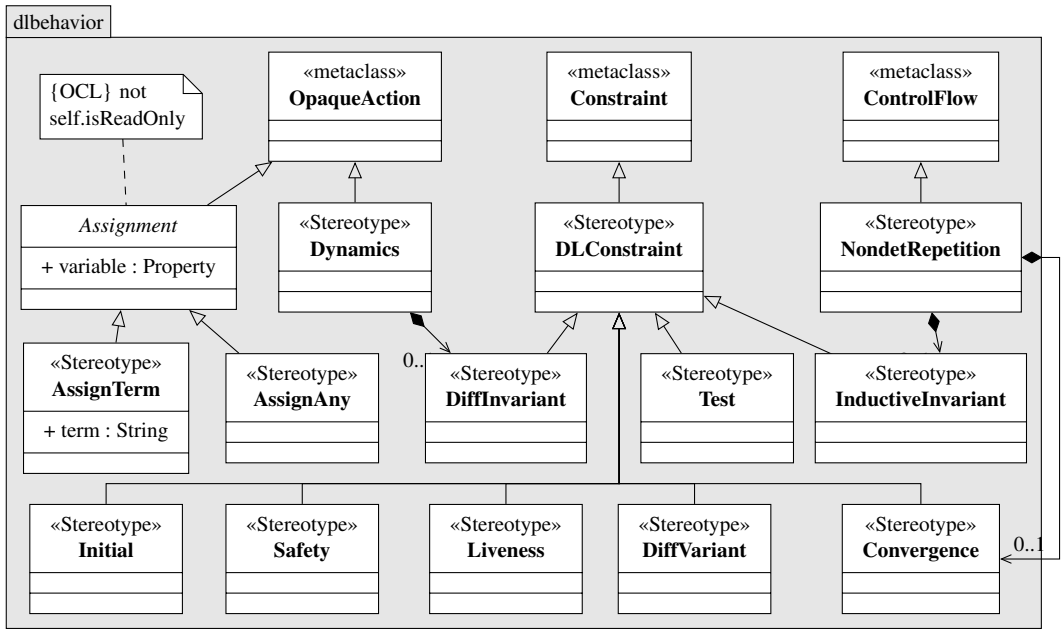

Hybrid programs know essentially two kinds of actions that can change the state of a system: instantaneous jumps (\ie, assignment) are part of the discrete control structure of a hybrid system, and differential equations are part of the continuous dynamics of a hybrid system.
\rref{fig:profile-dldynamics} shows the stereotypes for modeling the discrete and continuous dynamics of a hybrid system.

Activity diagrams, as introduced briefly above, provide modeling concepts to represent actions (opaque actions are essentially atomic blackbox actions), decisions, guarded control flow, and constraints.
%TODO Where does the dlbehavior diagram explain that assignment, test, dynamics, nondetrep are okay as statements but diffinv is not? 
In hybrid program UML we distinguish the following actions (actions are represented as rounded rectangles with a name compartment and a body compartment in UML): 
Nondeterministic assignment (\emph{AssignAny}) chooses any value for the variable.
Deterministic assignment (\emph{AssignTerm}) chooses the value defined by an arithmetic term for the variable.
Continuous evolution (\emph{Dynamics}) evolves the values of variables along a differential equation system that stays within a maximum evolution domain.
The differential invariant of the differential equation system, if known, can be annotated as a constraint to the dynamics action.
The common features of deterministic and nondeterministic assignment are factored into the abstract base class \emph{Assignment}.

The control flow of a hybrid program can be defined with three composition operations for hybrid program statements: nondeterministic choice, sequential composition, and nondeterministic repetition.
Nondeterministic choice can be modeled with standard UML notation for decisions (splitting and merging nodes), while sequential composition is control flow.
We introduce a stereotype \emph{NondetRepetition} for control flow (\ie, nondeterministic repetitions will be backwards edges), which can be annotated with a constraint to specify an inductive loop invariant.

Tests in hybrid programs ensure that a particular condition is satisfied in subsequent program statements.
They are modeled as algebraic constraints on control flows.
Further useful constraints are either part of the specification language (\emph{Initial}, \emph{Safety}, \emph{Liveness}) to express correctness criteria, or guide the verification process but do not influence system behavior (\emph{InductiveInvariant}, \emph{DiffInvariant}, \emph{Convergence}, \emph{DiffVariant}).

We define the semantics of Hybrid Program UML relative to \dL by transformation specifications from the Hybrid Program UML profile to the hybrid program metamodel.
Currently, we support two kinds of transformations: \emph{well-structured} activity diagrams\footnote{Well-structured activity diagrams consist of properly nested loops and branches and define a unique initial and final node.} can be transformed into well-structured hybrid programs with loops, whereas arbitrary activity diagrams can be transformed into \dL automata, which are hybrid automata embedded into hybrid programs with additional variables and tests to represent the states.

\paragraph{The Hybrid Program Semantics of Hybrid Program UML}
\label{sec:hpumlhpsemantics}

Currently, all constants and variables are handled globally as a flat structure (\ie, the structuring mechanisms present in a Hybrid Program UML model are not yet directly reflect in a hybrid program).
The hybrid programs $\alpha$ and $\beta$ in~\rref{tab:semantics-hp} represent either one of the atomic actions in Hybrid Program UML (assignment, nondeterministic assignment, or continuous dynamics) or a well-structured part of an activity diagram.
An edge between two actions \ovalbox{$\alpha$} and \ovalbox{$\beta$} corresponds to a sequential composition of the corresponding transformed hybrid programs $\alpha$ and $\beta$, cf.~\rref{rewrite:sequence}.
A guard on an edge is transformed into a sequential composition with an intermediate test, cf.~\rref{rewrite:testsequence}.
A decision node and a matching merge node with a forward edge and a backward edge is translated into a nondeterministic repetition, cf.~\rref{rewrite:repeat}.
If the forward edge is missing this means at least one repetition, cf.~\rref{rewrite:repeatleastonce}.
Decision nodes with tests are either translated into if-statements~\rref{rewrite:if} or if-else-statements~\rref{rewrite:ifelse}.
Finally, a decision node with a matching merge node (but without a back edge) is transformed into a nondeterministic choice, cf.~\rref{rewrite:choice}.
This transformation in~\rref{tab:semantics-hp} gives perfectly structured hybrid programs for well-nested activity diagrams.

\setcounter{tabline}{0}
\begin{table}[htb]
\caption{The hybrid program semantics of Hybrid Program UML}
\label{tab:semantics-hp}
\renewcommand{\arraystretch}{1.2}
\renewcommand{\tabularxcolumn}[1]{>{\normalsize}m{#1}}
\begin{tabularx}{\textwidth}{
  r
  m{3cm}
  m{2.5cm}
  X
  } 
\toprule
  & HP UML
  & Hybrid Program
  & Description
\tabularnewline \midrule 

(\tabrn)\label{rewrite:sequence} 
	& \begin{tikzpicture}
			[auto,
			location/.style ={rectangle, draw=black, thin, fill=white, text width=1em,align=center, rounded corners, minimum height=1.5em}
			]
			\node[location] (alpha) {$\alpha$};
			\node[location] (beta) [right of=alpha,xshift=1.5em] {$\beta$};
			\draw[style={->}] (alpha) -- (beta);
		\end{tikzpicture}
	& $\alpha;~\beta$ 
	& Direct control flow is a sequential composition\\
(\tabrn)\label{rewrite:testsequence}
	& \begin{tikzpicture}
			[auto,
			location/.style ={rectangle, draw=black, thin, fill=white, text width=1em,align=center, rounded corners, minimum height=1.5em}
			]
			\node[location] (alpha) {$\alpha$};
			\node[location] (beta) [right of=alpha,xshift=1.5em] {$\beta$};
			\draw[style={->}] (alpha) -- node{$\left[F\right]$} (beta);
		\end{tikzpicture} 
	& $\alpha;~\ptest{F};~\beta$ & Guarded control flow is a sequential composition with intermediate test\\
(\tabrn)\label{rewrite:repeat}
	& \begin{tikzpicture}
			[auto,
			location/.style ={rectangle, draw=black, thin, fill=white, text width=1em,align=center, rounded corners, minimum height=1.5em},
			decision/.style={diamond, draw=black, thin, fill=white, text width=0.1em,align=center, minimum height=0.1em}
			]
			\node[decision] (d1) {};
			\node[location] (alpha) [right of=d1] {$\alpha$};
			\node[decision] (d2) [right of=alpha] {};
			\draw[style={->}] (d1) -- (alpha);
			\draw[style={->}] (alpha) -- (d2);
			\draw[style={<->,rounded corners}] (d1) |- ($(alpha.south)+(0,-0.3)$) -| (d2);
		\end{tikzpicture}
	& $\alpha^*$ 
	& Decision node and merge node linked with backedge and forward edge is a nondeterministic repetition\\
(\tabrn)\label{rewrite:repeatleastonce} 
	& \begin{tikzpicture}
			[auto,
			location/.style ={rectangle, draw=black, thin, fill=white, text width=1em,align=center, rounded corners, minimum height=1.5em},
			decision/.style={diamond, draw=black, thin, fill=white, text width=0.1em,align=center, minimum height=0.1em}
			]
			\node[decision] (d1) {};
			\node[location] (alpha) [right of=d1] {$\alpha$};
			\node[decision] (d2) [right of=alpha] {};
			\draw[style={->}] (d1) -- (alpha);
			\draw[style={->}] (alpha) -- (d2);
			\draw[style={<-,rounded corners}] (d1) |- ($(alpha.south)+(0,-0.3)$) -| (d2);
		\end{tikzpicture}
	& $\alpha;~\alpha^*$ 
	& Decision node and merge node linked with backedge is at least one repetition\\
(\tabrn)\label{rewrite:if}
	& \begin{tikzpicture}
			[auto,
			location/.style ={rectangle, draw=black, thin, fill=white, text width=1em,align=center, rounded corners, minimum height=1.5em},
			decision/.style={diamond, draw=black, thin, fill=white, text width=0.1em,align=center, minimum height=0.1em}
			]
			\node[decision] (d1) {};
			\node[location] (alpha) [right of=d1,xshift=0.5em] {$\alpha$};
			\node[decision] (d2) [right of=alpha,xshift=-0.5em] {};
			\draw[style={->}] (d1) -- node{$\left[F\right]$} (alpha);
			\draw[style={->}] (alpha) -- (d2);
			\draw[style={->,rounded corners}] (d1) |- ($(alpha.south)+(0,-0.3)$) node[above,yshift=-0.5ex,pos=0.7]{$\left[\neg F\right]$} -| (d2);
		\end{tikzpicture}
	& $\text{if}(F)~ \alpha~ \text{fi}$ 
	& Decision node and merge node linked with forward edge is a conditional branch\\
(\tabrn)\label{rewrite:ifelse}
	& \begin{tikzpicture}
			[auto,
			location/.style ={rectangle, draw=black, thin, fill=white, text width=1em,align=center, rounded corners, minimum height=1.5em},
			decision/.style={diamond, draw=black, thin, fill=white, text width=0.1em,align=center, minimum height=0.1em}
			]
			\node[decision] (d1) {};
			\node[location] (alpha) [above right of=d1,xshift=0.5em,yshift=-0.6em] {$\alpha$};
			\node[location] (beta) [below right of=d1,xshift=0.5em,yshift=0.6em] {$\beta$};
			\node[decision] (d2) [right of=d1,xshift=2em] {};
			\draw[style={->,rounded corners}] (d1) |- node[pos=0.4]{$\left[F\right]$} (alpha);
			\draw[style={->,rounded corners}] (alpha) -| (d2);
			\draw[style={->,rounded corners}] (d1) |- node[below,pos=0.1,xshift=-1.3em]{$\left[\neg F\right]$} (beta);
			\draw[style={->,rounded corners}] (beta) -| (d2);
		\end{tikzpicture}
	& $\text{if}(F)~ \alpha~ \text{else}~ \beta~ \text{fi}$ 
	& Decision node and merge node with actions and mutually exclusive guards on each branch are if-else conditional\\
(\tabrn)\label{rewrite:choice} 
	& \begin{tikzpicture}
			[auto,
			location/.style ={rectangle, draw=black, thin, fill=white, text width=1em,align=center, rounded corners, minimum height=1.5em},
			decision/.style={diamond, draw=black, thin, fill=white, text width=0.1em,align=center, minimum height=0.1em}
			]
			\node[decision] (d1) {};
			\node[location] (alpha) [above right of=d1,yshift=-0.6em] {$\alpha$};
			\node[location] (beta) [below right of=d1,yshift=0.6em] {$\beta$};
			\node[decision] (d2) [right of=d1,xshift=1em] {};
			\draw[style={->,rounded corners}] (d1) |- (alpha);
			\draw[style={->,rounded corners}] (d1) |- (beta);
			\draw[style={->,rounded corners}] (alpha) -| (d2);
			\draw[style={->,rounded corners}] (beta) -| (d2);
		\end{tikzpicture}
	& $\alpha \cup \beta$ 
	& Decision node and matching merge node are a nondeterministic choice (analogous for more than two branches)
\tabularnewline \bottomrule
\end{tabularx}
\end{table}

\paragraph{The Hybrid Automaton Embedding Semantics of Hybrid Program UML}

The hybrid automaton embedding in hybrid programs defined in~\rref{tab:semantics-ha} matches smaller patterns in an activity diagram compared to the hybrid program transformation in~\rref{tab:semantics-hp}.
It is thus applicable to a wider range of activity diagrams, which do not even have to be well-structured (\ie, arbitrary state jumps are allowed).
As a downside, the transformation preserves no explicit program structure (\eg, sequence of statements) that could be exploited during verification.
This means that sufficient information about the program structure has to be conveyed in the system invariant.
This practice significantly increases the verification effort.

The hybrid automaton embedding constructs an automaton-structure from hybrid program notation instead of explicit program structure~\cite{Platzer10}.
It uses an additional  variable $s$ to keep track of the current location of the automaton.
A unique identifier per vertex and edge of the activity diagram identifies the automaton location.
The hybrid automaton embedding is then a nondeterministic choice over all the locations embedded in a single nondeterministic repetition\footnote{The final location does not change the location variable $s$. 
Thus, the system remains in the final location despite the fact that the nondeterministic repetition is allowed to execute arbitrarily many times.}, and the control flow is translated into updates of the current location with the respective follow-up location, as summarized in~\rref{tab:semantics-ha}.
The construction is analogous to the embedding of hybrid automata into hybrid programs~\cite[App. C]{Platzer10}, which follows exactly the same principles of implementing finite automata as programs.

\setcounter{tabline}{0}
\begin{table}[htb]
\caption{The hybrid automaton embedding semantics of Hybrid Program UML}
\label{tab:semantics-ha}
\renewcommand{\arraystretch}{1.2}
\renewcommand{\tabularxcolumn}[1]{>{\normalsize}m{#1}}
\begin{tabularx}{\textwidth}{
  r
  m{2cm}
  m{4cm}
  X
  } 
\toprule
  & HP UML
  & Hybrid Automaton Embedding in Hybrid Program
  & Description
\tabularnewline \midrule 

(\tabrn) 
	& \begin{tikzpicture}
			[auto,
			location/.style ={rectangle, draw=black, thin, fill=white, text width=1em,align=center, rounded corners, minimum height=1.5em},
			dloc/.style ={rectangle, draw=black, thin, dashed, fill=white, text width=1em,align=center, rounded corners, minimum height=1.5em},
			]
			\node[location] (alpha) {$\alpha$};
			\node[dloc] (beta) [right of=alpha] {$\beta$};
			\draw[style={->}] (alpha) -- (beta);
		\end{tikzpicture}
	& $\ptest{s = id(\alpha)};~ \alpha;~\humod{s}{id(\beta)}$ 
	& Action with direct control flow to another action is a sequential composition of a location test and the actual atomic action, with the transition modeled as assignment of a new location ID\\
(\tabrn) 
	& \begin{tikzpicture}
			[auto,
			location/.style ={rectangle, draw=black, thin, fill=white, text width=1em,align=center, rounded corners, minimum height=1.5em},
			dloc/.style ={rectangle, draw=black, thin, dashed, fill=white, text width=1em,align=center, rounded corners, minimum height=1.5em},
			decision/.style={diamond, draw=black, thin, fill=white, text width=0.1em,align=center, minimum height=0.1em},
			ddec/.style={diamond, draw=black, thin, dashed, fill=white, text width=0.1em,align=center, minimum height=0.1em}
			]
			\node[location] (alpha) {$\alpha$};
			\node[ddec] (decision) [right of=alpha] {};
			\draw[style={->}] (alpha) -- (decision);
		\end{tikzpicture}
	& $\ptest{s = id(\alpha)};~ \alpha;~\humod{s}{id(\Diamond)}$ 
	& Control flow between an action and a decision/merge node is a sequential composition of a state test and assignment of a new state ID\\
(\tabrn) 
	& \begin{tikzpicture}
			[auto,
			location/.style ={rectangle, draw=black, thin, fill=white, text width=1em,align=center, rounded corners, minimum height=1.5em},
			dloc/.style ={rectangle, draw=black, thin, dashed, fill=white, text width=1em,align=center, rounded corners, minimum height=1.5em},
			decision/.style={diamond, draw=black, thin, fill=white, text width=0.1em,align=center, minimum height=0.1em},
			ddec/.style={diamond, draw=black, thin, dashed, fill=white, text width=0.1em,align=center, minimum height=0.1em}
			]
			\node[location] (alpha) {$\alpha$};
			\node[dloc] (beta) [right of=alpha,xshift=0.5em] {$\beta$};
			\draw[style={->}] (alpha) -- node{$\left[F\right]$} (beta);
		\end{tikzpicture}
	& $\ptest{s=id(\alpha)};~\alpha;~\ptest{F};~\humod{s}{id(\beta)}$ 
	& Guarded control flow is a sequential composition with an intermediate test (analogous for decision/merge node)\\
(\tabrn) 
	& \begin{tikzpicture}
			[auto,
			location/.style ={rectangle, draw=black, thin, fill=white, text width=1em,align=center, rounded corners, minimum height=1.5em},
			decision/.style={diamond, draw=black, thin, fill=white, text width=0.1em,align=center, minimum height=0.1em},
			dloc/.style ={rectangle, draw=black, thin, dashed, fill=white, text width=1em,align=center, rounded corners, minimum height=1.5em},
			]
			\node[decision] (d1) {};
			\node[dloc] (alpha) [above right of=d1,yshift=-0.6em] {$\alpha$};
			\node[dloc] (beta) [below right of=d1,yshift=0.6em] {$\beta$};
			\draw[style={->,rounded corners}] (d1) |- (alpha);
			\draw[style={->,rounded corners}] (d1) |- (beta);
		\end{tikzpicture}
	& $\ptest{s=id(\Diamond)};~ \bigl(\humod{s}{id(\alpha)} \cup \humod{s}{id(\beta)}\bigr)$ 
	& Control flow between a decision node and actions is a nondeterministic choice (analogous for more than two branches)
\tabularnewline \bottomrule
\end{tabularx}
\end{table}

\subsubsection{Hybrid System Simulation}

An interesting opportunity for inspecting the behavior of a hybrid system during the modeling phase (prior to verification) is provided by Mathematica~9, which is able to simulate and plot hybrid system behavior using  a combination of \emph{NDSolve} and \emph{WhenEvent} conditions\footnote{\url{www.wolfram.com/mathematica}}. %\footnote{\url{www.wolfram.com/mathematica/new-in-9/advanced-hybrid-and-differential-algebraic-equations/}}. 
We transform corresponding excerpts of \dL to Mathematica for visualizing plots of the dynamic behavior of a hybrid program over time in \Sphinx.
Simulations can be useful for debugging system models and quickly conveying intuitions about their behavior to the respective members of the collaborative verification-driven engineering team.

%\todo{Challenges: (i) numerically solving ODEs: evolution domain is often slightly missed (\ie, no longer valid after NDSolve). We use heuristic to reset to maximum time in solution where evolution domain is still satisfied. Numeric errors also mean that we cannot use exact equality/inequality in tests. Instead, we use $\varepsilon$-equality/inequality (configurable). Together, these heuristics result in slightly unrealistic effects, such as a bouncing ball that bounces back slightly above ground. (ii) which values to use in nondeterministic assignment? (iii) which path to take in nondeterministic choice? (iv) how often execute nondeterministic repetition?}

\subsubsection{During the Proof}

\begin{figure}[tb]
%\centering
%        \begin{subfigure}[b]{\textwidth}
        	\includegraphics[width=.9\columnwidth]{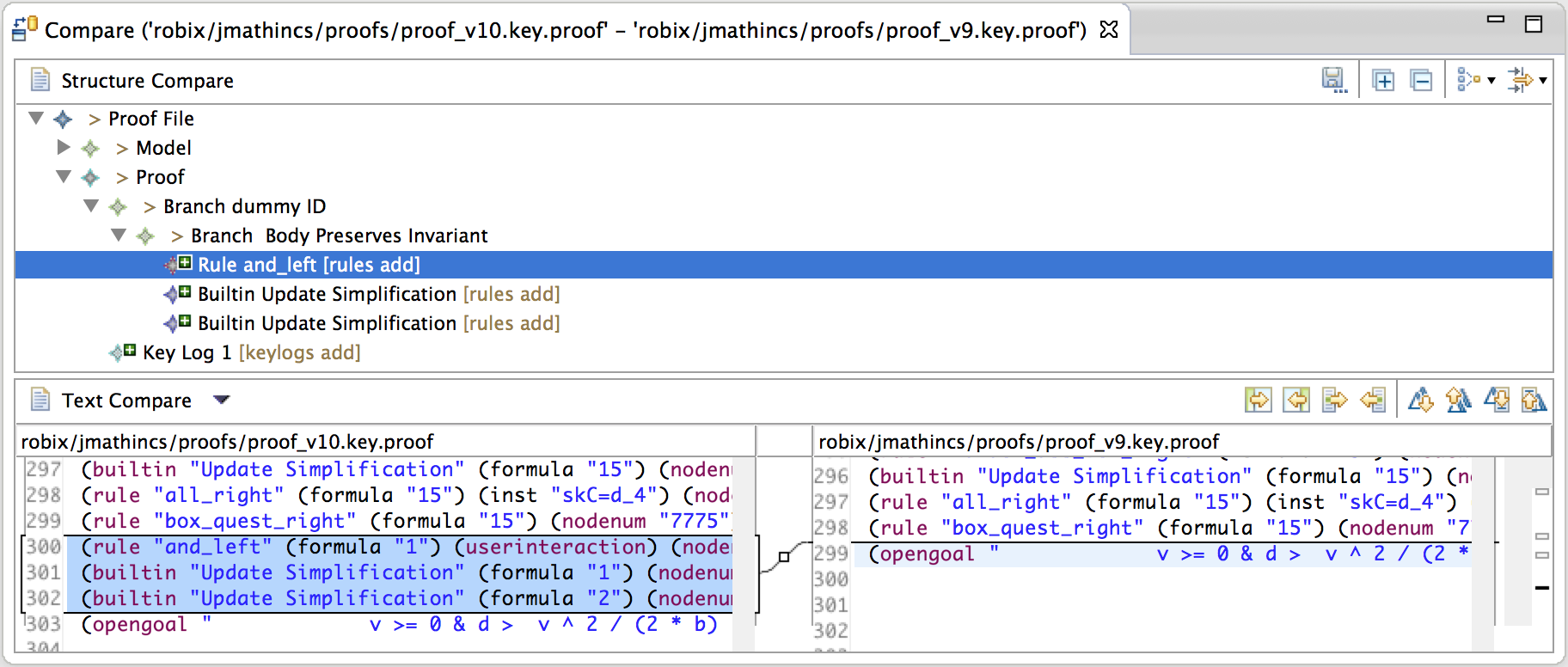}%{proofcomparison3.png}
%                \caption{Structural comparison}
%                \label{fig:proofcomparison_structure}
%        \end{subfigure}\\
%	        \begin{subfigure}[b]{\textwidth}
%        	\includegraphics[width=\columnwidth]{proofcomparisontextual.png}
%                \caption{Textual comparison}
%                \label{fig:proofcomparison_text}
%        \end{subfigure}%
\caption{Comparison of the structure of two proof versions}
\label{fig:proofcomparison}
\end{figure}

Collaboration support in \Sphinx includes model and proof comparison tools, both locally and with the model and proof repositories maintained in a central source code repository.
For this, not only textual comparison is implemented, but also structural comparison of models expressed in terms of the \dL metamodel and proofs expressed in terms of the \dL proof metamodel is supported (cf.~\rref{fig:proofcomparison}).
Exchanging proofs and inspecting updates on partial proofs is vital especially when multiple team members collaborate on a proof.
We inherit textual comparison integrated into our textual modeling editors from Eclipse and the Xtext framework.
Textual comparison of proofs, however, may not be the most efficient way of pointing to the relevant changes in a proof, because proofs, in addition to the relevant proof steps, often contain information for bookkeeping to mechanically check the proof.
Thus, we additionally inherit a structural comparison tool from the EMF Compare Diff/Merge framework\footnote{\url{http://www.eclipse.org/emf/compare/}} on the basis of a graph diff algorithm~\cite{Kolovos:2009:DMM:1564596.1564641}, so that differences are categorized by the relationships in the metamodel.
The relevant changes in the actual proof section of the proof file are then compared by their metamodel elements, which represent proof rules; if necessary, the additional bookkeeping code can be overlayed on demand.
\rref{fig:proofcomparison} shows a structural comparison overview in the top pane.
Four changes are in the proof section, which is expanded; three of them are located in the part of the inductive proof of a loop, where we have to show that the loop body preserves the invariant.
The first applied proof rule is selected and, therefore, the two lower panes show where this new proof step was inserted into the proof.

% Obtained by comparing proof_v9.key.proof with proof_v10.key.proof
% Switch to proof comparison (instead of model comparison)

Specific unsolved subproblems of a proof (\eg, complex arithmetic problems) can be flagged in \KeYmaera and extracted to other tools to further facilitate knowledge and expertise exchange.
That makes it easier to partition the verification effort and collaborate in jointly coming up with a solution. 
An open question, however, concerns the merging of the partial verification results into a single coherent proof without recourse to external verification steps.
In a first step, in \Sphinx we only allow exchanging proof strategies that can be executed by \KeYmaera.
\Sphinx injects these proof strategies directly into a \dL proof instead of an open goal, and \KeYmaera checks the injected proof steps for correctness.
That way, external (arithmetic) solvers can replace manual verification effort without compromising proof trustworthiness.

Later, actual proof certificates and further proof strategies will be exchanged to further increase trust, and more sophisticated comparisons of proof goals are envisioned to more robustly support replaying proofs.

\subsection{The Collaboration Backend}

\paragraph{Technical details.}
The \Sphinx modeling tool uses existing Eclipse plugins to connect to a variety of backend source code repositories and online project management  tools.
As source code repository we currently use Subversion\footnote{\url{subversion.apache.org}} and the Eclipse plugin Subclipse\footnote{\url{subclipse.tigris.org}}, but any other source code repository that is connected to Eclipse would work just as well.
Currently, Mylyn\footnote{\url{www.eclipse.org/mylyn}} and its connectors are used for accessing online project management tools (\eg, Bugzilla\footnote{\url{www.bugzilla.org}}, Redmine\footnote{\url{www.redmine.org}}, or any web-based tool via Mylyn's Generic Web templates connector) and exchanging tickets (\ie, requests for verification).
These tickets are the organizational means for collaborating on verification problems and tasks within a working group.
Exchange of models and proofs may then be conducted either by attaching files to tickets, or by linking tickets directly to models and proofs in the source code repository.
In the latter case, one benefits from the model and proof comparison capabilities of \Sphinx.
The collaboration through source code repositories and online project management systems seems useful because those are generally well-accepted by computer scientists and in other engineering domains; what still needs to be determined, however, is whether these collaboration tools are similarly well received by other domain experts.

\paragraph{Horizontal splitting.}
An important question for collaboration in cyber-physical systems verification is how domain experts with abstract, high-level understanding of a model can work with experts in control and dynamics to fill in details, and finally with verification experts to detect constraints that are necessary for verification.
Such a \emph{horizontal splitting} is in accordance with the distinction between high-level platform-independent models (PIM) and more detailed platform-specific models (PSM) as promoted by the OMG.
In our tool suite, domain experts can introduce placeholders into graphical models, which are essentially activity nodes with a descriptive name but without the formal definition of their underlying meaning (\eg, the dynamics of the water tank as a differential equation).
\Sphinx translates placeholders into comments in the formal language, so that other experts (\eg, in modeling motion with differential equations) can fill in those gaps.
Moreover, if \Sphinx encounters a placeholder in a graphical model, it augments the safety condition with $\textit{false}$ so that all proof attempts on incomplete models will fail.
Domain experts can even omit tests (constraints) on transitions.
The absence of constraints that are essential for safety will be detected by the theorem prover \KeYmaera anyway, as described below.

\paragraph{Constraint detection.}
Often, a development team has a good overall understanding of the desired safety conditions, but not of all of the corner cases and constraints that make the system safe \wrt these safety conditions.
Therefore, one of the most challenging problems in cyber-physical systems verification is to find such constraints (\eg, loop invariants or switching constraints).
\KeYmaera can help finding constraints in various ways, as done in a prior case study on train control~\cite{DBLP:conf/icfem/PlatzerQ09}.
\KeYmaera syntactically decomposes hybrid programs, so that only arithmetic proof obligations remain towards the leaves of a proof.
These arithmetic proof obligations, if being unprovable, reveal specific valuations as counterexamples and the assumptions that were collected up to that point.
\begin{itemize}
\item The path leading to an unprovable goal encodes the specific location in the model; it may thus guide a domain expert to an important corner case in the model.
\item A counterexample may reveal necessary initial and invariant conditions when choosing an uncontrolled dynamics model~\cite{DBLP:conf/icfem/PlatzerQ09} as input to \KeYmaera.
\end{itemize}
The discovered constraints can then be communicated to the domain expert for review before the verification continues.

\section{Application Example}
\label{sec:application}

In this section we illustrate a verification example of an autonomous robot~\cite{DBLP:conf/rss/MitschGP13} that we collaboratively developed and solved using \KeYmaera and geometric relevance filtering~\cite{ppz:grf:2013}.
We compare the effort of using \KeYmaera interactively, \KeYmaera fully automated, and \KeYmaera together with geometric relevance filtering connected via \Sphinx.

With the increased introduction of autonomous robotic ground vehicles as consumer products---such as autonomous hovers and lawn mowers, or even accepting driverless cars on regular roads in California---we face an increased need for ensuring product safety not only for the good of our consumers, but also for the sake of managing manufacturer liability.
One important aspect in building such systems is to make them scrutable, in order to mitigate unrealistic expectations and increase trust~\cite{Tintarev2013}.
In the design stage of such systems, formal verification techniques ensure correct functioning \wrt some safety condition, and thus, increase trust.
In the course of this, formal verification techniques can help to make assumptions explicit and thus clearly define what can be expected from the system under which circumstances (before the system is built and executed).

We are going to illustrate a design and verification process that encourages collaboration from high-level graphical models which convey intuition about the system to a broad and possibly heterogeneous audience to detailed formal models, which are suitable for formal verification.
For this we will discuss our formal model of an autonomous robotic ground vehicle and its proof.
More details on the model and case studies, as well as extensions for moving obstacles, sensor uncertainty, sensor failure, and actuator disturbance can be found in~\cite{DBLP:conf/rss/MitschGP13}.

We will begin with a hierarchically structured graphical model that defines the high-level system behavior, the expected operating environment and the initial conditions under which the robot can be activated safely together with the invariants and safety conditions that the robot will then guarantee.
We will complement the high-level model with a more detailed robot controller model.
Together, these models are translated into \dL and formally verified.
Finally, we will discuss a simple one-dimensional model of the robot to exemplify how modeling decisions in \dL can make verification easier.

\subsection{Hierarchical Graphical Modeling}

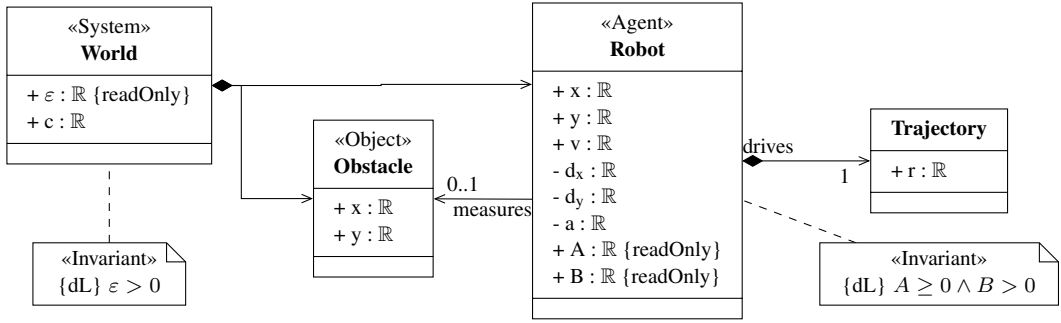
\begin{figure}[htb]
\centering
\begin{footnotesize}
\begin{tikzpicture}
\umlclass[type=System,y=1]{World}
	{+ $\varepsilon$ : $\mathbb{R}$ \{readOnly\}\\
	 + c : $\mathbb{R}$	
	}
	{}
\umlclass[type=Agent,x=7]{Robot}
	{+ x : $\mathbb{R}$	\\
	 + y : $\mathbb{R}$	\\
	 + v : $\mathbb{R}$	\\
	 - $\text{d}_\text{x}$ : $\mathbb{R}$\\
	 - $\text{d}_\text{y}$ : $\mathbb{R}$\\
	 - a : $\mathbb{R}$\\
	 + A : $\mathbb{R}$ \{readOnly\}\\
	 + B : $\mathbb{R}$	 \{readOnly\}\\
	}
	{}
\umlclass[x=11]{Trajectory}
	{+ r : $\mathbb{R}$
	}
	{}
%\umlclass[type=Object,x=5,y=-4]{Obstacle}
%	{+ x : $\mathbb{R}$	\\
%	 + y : $\mathbb{R}$	\\
%	}
%	{}
\umlclass[type=Object,x=3.5,y=-0.5]{Obstacle}
	{+ x : $\mathbb{R}$	\\
	 + y : $\mathbb{R}$	\\
	}
	{}

\umlunicompo[geometry=-|-,arm2=-2cm,anchor2=144]{World}{Robot}
\umlunicompo[geometry=-|-]{World}{Obstacle}
\umlunicompo[arg1=drives,mult=1]{Robot}{Trajectory}
\umluniassoc[arg1=measures,pos1=0.4,mult=0..1,pos2=0.7,anchor1=200]{Robot}{Obstacle}

\umlnote[x=11,y=-1.5,width=23ex]{Robot}{\vspace{-2ex}\begin{center}\UmlSt{Invariant}\\\{dL\} $A \geq 0 \wedge B > 0$\end{center}}
\umlnote[y=-1.5,width=14ex]{World}{\vspace{-2ex}\begin{center}\UmlSt{Invariant}\\\{dL\} $\varepsilon > 0$\end{center}}
\end{tikzpicture}
\end{footnotesize}
\caption{The structure of the robotic obstacle avoidance model}
\label{fig:eval-structure}
\end{figure}

First, we construct a high-level model of the structure and the behavior of the robotic ground vehicle and of the assumptions about the environment it is operating in.
\rref{fig:eval-structure} uses the Hybrid Program UML profile to model the structure of the robotic obstacle avoidance algorithm.

The system class $\textit{World}$ provides a global clock $c$ and ensures a cycle time of at most $\varepsilon$ time units (\ie, any controller in the system will run at least once every $\varepsilon$ time units).
The state of a robot is characterized by its position ($x$, $y$) and orientation ($d_x$, $d_y$) in two dimensions and its linear velocity ($v$). 
The robot can control its linear acceleration within certain bounds ($a \in \left[-B,A\right]$) and choose a new trajectory.
It measures the position of the nearest obstacle to make decisions about its trajectory.

\begin{figure}[htb]
\centering
\begin{footnotesize}
\begin{tikzpicture}
\umlstateinitial[name=initial,y=0.5]
\umlstatedecision[name=loopbegin,x=1,y=0.5]
%\begin{umlstate}[name=ctrl,x=3]{ctrl}

%\end{umlstate}
\umlbasicstate[name=ctrl,x=3,type=X,do=\protect{see \rref{fig:eval-ctrlbehavior}}]{ctrl}
%\umlbasicstate[name=resetclock,x=6,do=$\humod{c}{0}$]{reset clock}
\umlbasicstate[name=dyn,x=7.5,width=30ex,do=\protect{$\humod{c}{0};~ \left(\D{x} = v d_x \ldots ~\&~ c \leq \varepsilon\right)$}]{\UmlSt{Dynamics} dyn}
\umlstatedecision[name=loopend,x=11,y=0.5]
\umlstatefinal[name=final,x=12,y=0.5]

\umltrans{initial}{loopbegin}
\umltrans{loopbegin}{ctrl}
%\umltrans{ctrl}{resetclock}
%\umltrans{resetclock}{dyn}
\umltrans{ctrl}{dyn}
\umltrans{dyn}{loopend}
\umltrans{loopend}{final}
\umlrelation[style=<->,geometry=|-|, arm1=-1.3cm,name=loop,stereo=NondetRepetition,pos stereo=1.5]{loopend}{loopbegin}
%\umlVHVtrans[arm1=-0.8cm,name=loop,stereo=NondetRepetition,pos stereo=1.5]{loopend}{loopbegin}

\umlnote[x=1.5,y=-2,width=40ex]{initial}{\vspace{-2ex}\begin{center}\UmlSt{Initial}\\\{dL\} $A \geq 0 \land B > 0 \land \varepsilon > 0$\\ $\land v \geq 0 \land \lVert p_r - p_o \rVert_\infty > \frac{v^2}{2 B}$\end{center}}
\umlnote[x=11,y=-2,width=23ex]{final}{\vspace{-2ex}\begin{center}\UmlSt{Safety}\\\{dL\} $\lVert p_r - p_o \rVert_2 > 0$\end{center}}
\umlnote[x=6.8,y=-2,width=35ex]{loop-3}{\vspace{-2ex}\begin{center}\UmlSt{Invariant}\\\{dL\} $v \geq 0 \land \lVert p_r - p_o \rVert_\infty > \frac{v^2}{2 B}$\end{center}}
\end{tikzpicture}
\end{footnotesize}
\caption{Overview of the behavior of the robotic obstacle avoidance model.
The model is structured hierarchically, with details on $\textit{ctrl}$ specified in~\rref{fig:eval-ctrlbehavior}}
\label{fig:eval-systembehavior}
\end{figure}
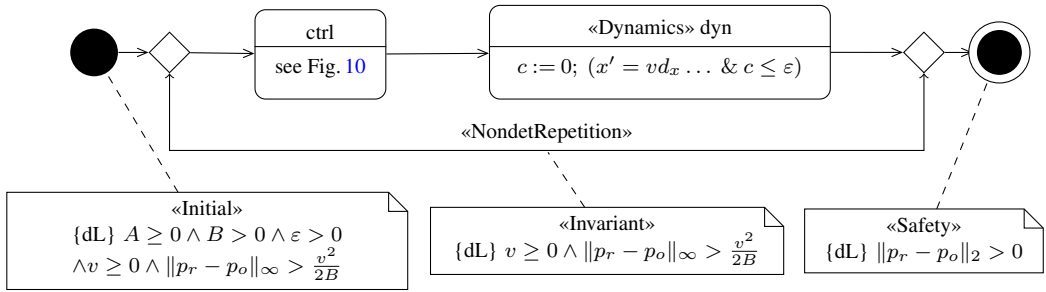

The high-level behavior of the robotic obstacle avoidance algorithm is modeled  in a hierarchical activity diagram using our Hybrid Program UML profile.
\rref{fig:eval-systembehavior} shows the high-level behavior with the controller and the dynamics.
In this example, the dynamics is a non-linear differential-algebraic equation that describes the robot's motion on a circular segment: $\D{x}=v d_x, \D{y}=v d_y, \D{d_x}=-\frac{v d_y}{r}, \D{d_y}=\frac{v d_x}{r}, \D{v}=a ~\&~ v \geq 0 \land c \leq \varepsilon$.
The high-level behavior further details the initial condition under which the obstacle avoidance algorithm is safe to start and the safety condition that we want to be true for all executions (in these conditions we use $p_r=(x,y)$ to denote the position of the robot and $p_o$ to denote the position of the obstacle in two dimensions).

%$\D{x_r}=v_r dx_r,\D{y_r}=v_r dy_r, \D{dx_r}=-\frac{a_r dy_r}{r}, \D{dy_r}=\frac{a_r dx_r}{r}, \D{v_r}=a_r,\D{c}=1 ~\&~ c \leq \varepsilon \wedge v_r \geq 0$

\begin{figure}[htb]
\centering
\begin{footnotesize}
\begin{tikzpicture}
\umlstateinitial[name=initial,y=0.5]
\umlstatedecision[name=choicebegin,x=1,y=0.5]
\umlbasicstate[name=sense,x=3,y=1.5,do=$\humod{p_o}{*}$]{\UmlSt{AssignAny} sense}
\umlbasicstate[name=brake,x=6,do=$\humod{a}{-B}$]{\UmlSt{AssignTerm} brake}
\umlbasicstate[name=coast,x=6,y=-1.5,do=$\humod{a}{0}$]{\UmlSt{AssignTerm} stop}

\umlbasicstate[name=traj,x=6,y=1.5,do=$\humod{r}{*}$]{\UmlSt{AssignAny} curve}
\umlbasicstate[name=acc,x=9,y=1.5,do=$\humod{a}{*}$]{\UmlSt{AssignAny} acc}

\umlstatedecision[name=choiceend,x=11,y=0.5]

\umlstatefinal[name=final,x=12,y=0.5]

\umltrans{initial}{choicebegin}
\umlVHtrans[name=sense]{choicebegin}{sense}
\umltrans{choicebegin}{brake}
\umlVHtrans[name=stopped,anchor2=180]{choicebegin}{coast}
\umltrans{sense}{traj}
\umltrans{traj}{acc}
%\umlVHVtrans[arm1=-1.8cm,name=safe]{acc}{choiceend}
\umlHVtrans[name=safe]{acc}{choiceend}
\umltrans{brake}{choiceend}
\umlHVtrans{coast}{choiceend}
\umltrans{choiceend}{final}

\umlnote[x=3,y=-0.2,width=15ex]{stopped-3}{\vspace{-2ex}\begin{center}\UmlSt{Test}\\\{dL\} $\ptest{v=0}$\end{center}}

\umlnote[x=1.8,y=3.5,width=50ex]{sense}{\vspace{-2ex}\begin{center}\UmlSt{Test}\\\{dL\}
$\lVert p_r - p_o \rVert_\infty > \frac{v^2}{2 B} + \left(\frac{A}{B} + 1\right)\left(\frac{A}{2}\varepsilon^2 + \varepsilon v\right)$\end{center}}

\umlnote[x=6.5,y=3.5,width=13ex]{traj}{\vspace{-2ex}\begin{center}\UmlSt{Test}\\\{dL\} $r \neq 0$\end{center}}

\umlnote[x=9.5,y=3.5,width=20ex]{acc}{\vspace{-2ex}\begin{center}\UmlSt{Test}\\\{dL\} $-B \leq a \leq A$\end{center}}

%\umlnote[x=12,y=2.2]{safe-3}{$r \neq 0 \land -B \leq a \leq A \land \lVert p_r - p_o \rVert_\infty > \frac{v^2}{2 B} + \left(\frac{A}{B} + 1\right)\left(\frac{A}{2}\varepsilon^2 + \varepsilon v\right)$}
\end{tikzpicture}
\end{footnotesize}
\caption{The controller of the robotic obstacle avoidance model}
\label{fig:eval-ctrlbehavior}
\end{figure}
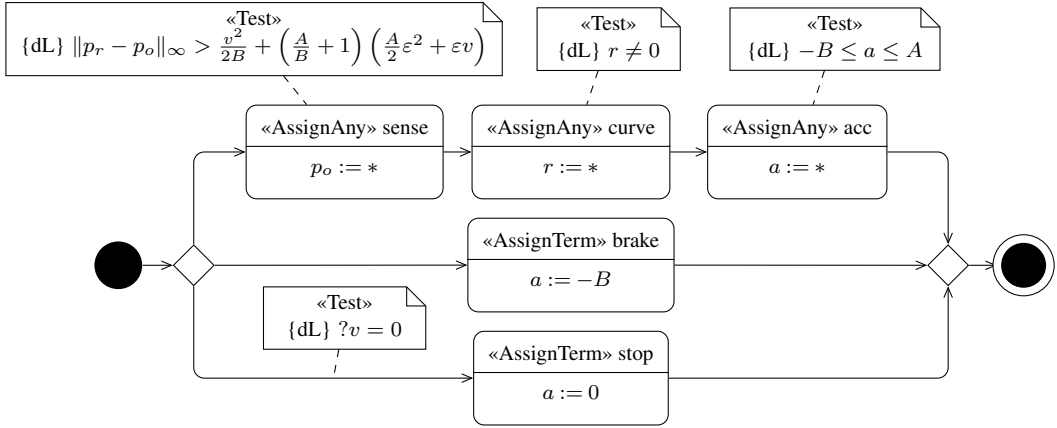

A model of such high-level behavior is useful to communicate major design decisions, such as the expected operating environment and the most important constraints that the system must obey.
It also consolidates more detailed models that may have been produced by different members of a verification team.
As future work we will integrate composite structure diagrams, as in~\cite{DBLP:journals/sttt/BerkenkotterBHP06}, to make the interfaces between those detailed models explicit.
A more detailed model of the controller complements the high-level $\textit{ctrl}$ block with detailed implementation-specific decisions, as shown in~\rref{fig:eval-ctrlbehavior}.

The robot has three control options (top to bottom in~\rref{fig:eval-ctrlbehavior}): 
If the robot's current state is safe with respect to the sensed position of the nearest obstacle, then the robot may choose a new curve and accelerate with any rate within its physical bounds.
For this, we utilize the modeling pattern introduced above: we assign an arbitrary value to the robot's acceleration state ($\humod{a}{*}$), which is then restricted to any value from the interval $\left[-B,A\right]$ using a test ($\ptest{-B\leq a \leq A}$).
The robot can brake ($\humod{a}{-B}$), which we want to be an emergency action that should be executed with minimal time delay (\ie, we want braking to be safe even when the robot relies on previously sensed obstacle positions).
Finally, if the robot is stopped ($\ptest{v=0}$), it may choose to remain in its current spot ($\humod{a}{0}$).

To stay always safe the robot must account for (i) its own braking distance ($\tfrac{v^2}{2 B}$), (ii) the distance it may travel with its current velocity ($\varepsilon v$) until it is able to initiate braking, and (iii) the distance needed to compensate the acceleration $A$ that may have been chosen in the worst case.
For a complete model of the robotic obstacle avoidance algorithm and further variants as a hybrid program we refer to~\cite{DBLP:conf/rss/MitschGP13}.

\subsection{The Effect of Collaboration on Arithmetic Verification Effort}

In the following paragraphs we discuss how the structure of the robotic obstacle avoidance algorithm and the resulting proof structure can be exploited to facilitate collaboration during the proof.
Furthermore, we describe variants of the proof with varying degree of manual guidance and with/without collaboration using geometric relevance filtering~\cite{ppz:grf:2013}.
This way, we are able to give a comparison on the proof effort that is necessary to discharge arithmetic proof obligations with and without collaboration. 

The \dL proof calculus provides proof rules to syntactically decompose a hybrid program into smaller, easier provable pieces.
Such a proof unfolds into many subgoals that often can be handled separately.
\rref{proof:robot} sketches the proof structure of the robot obstacle avoidance safety proof together with the proof rules used in the proof sketch\footnote{The \dL proof calculus is explained in detail in~\cite{Platzer10,DBLP:conf/lics/Platzer12b}}.
The names in the proof are the abbreviations that we introduced in the graphical model as placeholders for more complicated formulas, which get expanded when necessary.
The three control choices of $\textit{ctrl}$ are transformed by the proof rule $\irref{choicebr}$ into a conjunction, which is further split by the proof rule $\irref{andr}$ into separate branches in the proof.

\begin{proof}[h!]
\begin{footnotesize}
 \begin{calculuscollections}{\textwidth}
   \begin{calculuscollection}
    \begin{calculus}[context=L]
		\cinferenceRule[choicebr|$\dibox{\cup}\rightrule$]{}
		{\linferenceRule
		  {\lsequent{}{\dbox{\alpha}{\phi}\land\dbox{\beta}{\phi}}}
		  {\lsequent{}{\dbox{\pchoice{\alpha}{\beta}}{\phi}}}
		}{}
	\end{calculus}    
  \end{calculuscollection}
  \hspace{\linferenceRulehskipamount}\hspace{0.4cm}     
  \begin{calculuscollection}[prefix=P,reset,context=L]
    \linferenceRulehskipamount=4mm%
    \begin{calculus}
      \cinferenceRule[andr|$\land$\rightrule]{$\land$ right}
      {\linferenceRule[sequent]
        {\lsequent{}{\phi}
          & \lsequent{}{\psi}}
        {\lsequent{}{\phi \land \psi}}
      }{}
    \end{calculus}
  \end{calculuscollection}
  \hspace{\linferenceRulehskipamount}\hspace{0.4cm}     
  \begin{calculuscollection}[prefix=D,reset]
    \begin{calculus}
      \cinferenceRule[composeb|$\dibox{{;}}$]{composition}
      {\linferenceRule[sequent]
        {\lsequent[s]{}{
              \dbox{\alpha}{\dbox{\beta}{\phi}}}}
        {\lsequent[s]{}{
              \dbox{\alpha;\beta}{\phi}}}
      }{}
    \end{calculus}
  \end{calculuscollection}%
  \hspace{\linferenceRulehskipamount}\hspace{0.4cm}     
%%  \begin{calculuscollection}[prefix=D,reset]
%%  \begin{calculus}
%%  \cinferenceRule[testb|$\dibox{?}$]{test}
%%      {\linferenceRule[sequent]
%%        {\lsequent[s]{}{\ivr \limply \psi}}
%%        {\lsequent[s]{}{\dbox{\ptest{\ivr}}{\psi}}}
%%      }{}
%%  \end{calculus}
%%  \end{calculuscollection}
  \hspace{\linferenceRulehskipamount}\hspace{0.4cm}
  \begin{calculuscollection}
    \begin{calculus}
      \cinferenceRule[weakenr|W\rightrule]{weakening right}
      {\linferenceRule[sequent]
        {\lsequent{}{}}
        {\lsequent{}{\phi}}
      }{}
    \end{calculus}
   \end{calculuscollection}
   \hspace{\linferenceRulehskipamount}\hspace{0.4cm}
   \begin{calculuscollection}
    \begin{calculus}
 	 \cinferenceRule[weakenl|W\leftrule]{weakening left}
      {\linferenceRule[sequent]
        {\lsequent{}{}}
        {\lsequent{\phi}{}}
      }{}
      \irlabel{qe|\text{QE}}
      \irlabel{deva|\text{expert A}}
      \irlabel{devb|\text{B}}
      \irlabel{devc|\text{C}}
    \end{calculus}
  \end{calculuscollection}
  \end{calculuscollections}

%\vspace{2ex}
\begin{sequentdeduction}[default]
	\linfer[choicebr+andr]
	{
	\linfer[qe]
	{\lclose}
	{
	\linfer[weakenl+weakenr]
	{\lsequent{\tilde{\phi}}{\tilde{\psi}}}
	{
	\linfer[deva]
	{\lsequent{\phi \ldots}{\psi \ldots}}
	{
	\lsequent{\phi}{\dibox{\textit{sense};~\textit{curve};~\textit{acc}}\dibox{\textit{dyn}}\psi}
	}%deva
	}%weaken
	}%qe
	&
	\linfer[choicebr+andr]
	{
	\linfer[devb]
	{\ldots}
	{\lsequent{\phi}{\dibox{\textit{brake}}\dibox{\textit{dyn}}\psi}
	}%devb
	 &
	\linfer[devc]
	{\ldots}
	 {
	 \lsequent{\phi}{\dibox{\textit{\ptest{v=0};~\textit{stop}}}\dibox{\textit{dyn}}\psi}
	 }%devc
	}
	{	
	\lsequent{\phi}{\dibox{(\textit{brake}) \cup (\ptest{v=0};~\textit{stop})}\dibox{\textit{dyn}}\psi}}
	}%choicebr+andr
	{
	\linfer[]
	{\lsequent{\phi}{\dibox{(\textit{sense};~\textit{curve};~\textit{acc}) \cup (\textit{brake}) \cup (\ptest{v=0};~\textit{stop})} \dibox{\textit{dyn}} \psi}}
	{
	\linfer[composeb]
	{\lsequent{\phi}{\dibox{\textit{ctrl}} \dibox{\textit{dyn}} \psi}}
	{\lsequent{\phi}{\dibox{\textit{ctrl};~\textit{dyn}}\psi}
	}%composeb
	}%expand	
	}%choicebr+andr
 \end{sequentdeduction} 
\end{footnotesize}
\caption{Proof sketch of the robot obstacle avoidance algorithm using indicated proof rules}\label{proof:robot}
\end{proof}

These branches can be handled separately by different verification team members, who apply further proof rules of the \dL proof calculus to continue the proof (cf. branches \irref{deva}, \irref{devb} and \irref{devc}).
Towards the leaves of a branch the proof rules of \dL increasingly eliminate hybrid program elements by turning them into first-order real arithmetic formulas.
These formulas are often hard to prove, because the \dL proof rules are not designed to automatically identify and eliminate unnecessary context information (\eg, $\phi$ still contains information about acceleration, even though it is irrelevant to prove braking safety).
Quantifier elimination, which is the final step to proof correctness of a first-order real arithmetic formula, is doubly exponential in the formula size~\cite{Davenport1988}.
This means that we want to reduce the number of variables at the leaves of the proof as much as possible.
At this stage collaboration across different verification tools becomes possible: we can ship off the formulas to an arithmetic tool or expert to discover what information is unnecessary and can then weaken the formulas in the sequent (\irref{weakenl},\irref{weakenr}) before we invoke the quantifier elimination procedure (\irref{qe}).

We compare different proof variants of the robot obstacle avoidance algorithm to highlight the potential reduction in proof effort when developers with different expertise collaborate on a proof.
\rref{tab:proofcomplexity} compares the number of proof branches, the total number of proof steps, the number of manually executed steps, the number of manually executed weaken operations, the number of exported goals and goals solved by the external tool, the proof execution duration, and the memory used during the arithmetic in the proof.
The baseline (line 1 in~\rref{tab:proofcomplexity}) is a proof with manual optimization to reduce branching.
We created two further variants of the proof: the guided variant manually weakened obviously unnecessary contextual information to reduce the number of branches in the proof; the mechanic variant branches fully automated by \KeYmaera.
Both variants were finished fully interactively (cf. lines 2 and 5  in~\rref{tab:proofcomplexity}), fully automated in \KeYmaera (cf. lines 3 and 6 in~\rref{tab:proofcomplexity}), and automated with real-arithmetic formulas exported to geometric relevance filtering (cf. lines 4 and 7 in~\rref{tab:proofcomplexity}). 

\begin{table}[htb]
\caption{Proof effort in \KeYmaera with and without collaboration}
\label{tab:proofcomplexity}
\begin{tabularx}{\linewidth}{
  r
  X
  r
  r
  r
  r
  r
  r
  r
  } 
\toprule
  & Variant
  & \multicolumn{2}{c}{Proof Size}
  & \multicolumn{2}{c}{Manual Steps}
  & Exported
  & Time [s]
  & Mem.
  \tabularnewline \cmidrule(r){3-4}\cmidrule{5-6}
  &
  & \makebox[0.3\width][r]{Branches}
  & Steps
  & All
  & \irref{weakenl},\irref{weakenr}
  & \raisebox{1.5ex}[0pt]{(Solved)}
  & Full (Arith.)
  & \raisebox{1.5ex}[0pt]{[MB]}
  \tabularnewline \midrule 
(1) & Baseline & 67 & 868 & 139 & 84  & 0 & 34.8 (2.4) & 51.7\\
(2) & Guided (interactive finish) & 67 & 935 & 202 & 148 & 0 & 45.8 (11.6) & 52.9\\
(3) & Guided (auto finish) & \multicolumn{7}{c}{aborted after >2\,h}\\
(4) & Guided (GRF finish) & 67 & 980 & 108 & 54 & 18 (13) & 38.1 (4) & 52.2 \\
(5) & Mechanic (interactive finish) & 87 & 1193 & 440 & 356 & 0 & 46.9 (3.7) & 52.2 \\
(6) & Mechanic (auto finish) & \multicolumn{7}{c}{aborted after >2\,h}\\
(7) & Mechanic (GRF finish) & 87 & 1230 & 139 & 55 & 32 (28) & 46.4 (4.2) & 52.4
  \tabularnewline \bottomrule
\end{tabularx}
\end{table}

The interesting result is that geometric relevance filtering can solve many of the cases introduced by the fully mechanic branching, while it fails on the same highly complex problems as in the partly mechanic case.
This means that the external tool directs the manual effort that is still needed in both variants to the interesting cases, while it takes care of much of the tedious work.
Thus, although less manual effort was put into guiding the automated branching of \KeYmaera, the effort for reducing arithmetic goals to a manageable size for quantifier elimination procedures was reduced by prior collaboration involving geometric relevance filtering~\cite{ppz:grf:2013}.

\subsection{The Effect of Model Variants on Proof Structure}
\label{sec:variants}

Since it is hard to come up with a fully verifiable model that includes all the details right from the beginning, the models discussed in the previous section and in our previous case studies~\cite{DBLP:conf/iccps/MitschLP12,DBLP:conf/do-form/MitschPP13, DBLP:conf/rss/MitschGP13} are the result of different modeling and verification variants.
In the process of creating these models, different assumptions and simplifications were applied until we reached the final versions.
We developed proof-aware refactoring methods to carry over verified properties about an original model to a refactored model~\cite{DBLP:conf/fm/MitschQP14}, in order to reduce proof effort.

In this section, we discuss how various design decisions influence the structure of a proof and, in turn, the verification effort.

\subsubsection{Modeling}

We use a simplified model of a robot on a one-dimensional track~\cite{DBLP:conf/do-form/MitschPP13}.
In this example, navigation of a robot is considered safe, if the robot is able to stay within its assigned area (\eg, on a track) and does not actively crash with obstacles.
Since we cannot guarantee reasonable behavior of obstacles, however, the robot is allowed to passively crash (\ie, while obstacles might run into the robot, the robot will never move into a position where the obstacle could not avoid a collision).
%
%
% Model: robix/map/03d-robot_1D-MovingObstacle-NoDestination-SimpleNav.key
%
\begin{model}[b!]
\caption{Single wheel drive without steering (one-dimensional robot navigation)}
\label{model:1d}
\begin{footnotesize}
\begin{align}
\label{eq:03:1}\text{\it swd}  & \equiv (\text{\it ctrl};\text{\it dyn})^*\\
\label{eq:03:2}\text{\it ctrl} & \equiv(\text{\it ctrl}_r \parallel \text{\it ctrl}_o)\\
\label{eq:03:2-1}\text{\it ctrl}_r & \equiv(a_r := -B)\\
\label{eq:03:2-2}			& \phantom{\equiv}\cup (?\text{\it safe};~ a_r := *;~ ?-B \leq a_r \leq A)\\
\label{eq:03:2-3}		  & \phantom{\equiv}\cup (?v_r = 0;~ a_r := 0;~ o_r := *;~ ?o_r^2=1)\\
\label{eq:03:2-5}\text{\it safe} & \equiv x_{\underline{b}} + \frac{1-o_r}{2}\Biggl(\frac{v_r^2}{2 B} + \left(\frac{A}{B} + 1\right) \left(\frac{A}{2} \varepsilon^2 + \varepsilon v_r\right)\Biggr) < x_r <   
    x_{\overline{b}} - \frac{1+o_r}{2}\Biggl(\frac{v_r^2}{2 B} + \left(\frac{A}{b} + 1\right) \left(\frac{A}{2} \varepsilon^2 + \varepsilon v_r\right)\Biggr)\\
\label{eq:03:2-6}		& \phantom{\equiv} \wedge \norm{x_r - x_o} \geq \frac{v_r^2}{2 B}+\left(\frac{A}{B}+1\right)\left(\frac{A}{2} \varepsilon^2+\varepsilon v_r\right) + V\left(\varepsilon+\frac{v_r+A \varepsilon}{B}\right)\\ 
\label{eq:03:3-1}\text{\it ctrl}_o & \equiv\left(?v_o=0;~ o_o := *;~ ?o_o^2 = 1\right)\\
\label{eq:03:3-2}		&\phantom{\equiv} \cup (v_o:=*;~ ?0 \leq v_o \leq V)\\
\label{eq:03:7}\text{\it dyn} & \equiv (t := 0;~ x_r'=o_r v_r,~ v_r'=a_r,~ x_o'=o_o v_o,~ t'=1~ \&~ v_r \geq 0 \wedge v_o\geq 0 \wedge t \leq \varepsilon)
\end{align}
\end{footnotesize}
\end{model}
\rref{model:1d} shows a textual \dL model of a hybrid system comprising the control choices of an autonomous robotic ground vehicle, the control choices of a moving obstacle, and the continuous dynamics of the system.
The system represents the common controller-plant model: it repeatedly executes control choices followed by dynamics, cf.~\eqref{eq:03:1}.
The control of the robot is executed in parallel to that of the obstacle, cf.~\eqref{eq:03:2}.

Once again, the robot has three options: it can brake unconditionally, cf.~\eqref{eq:03:2-1}.
If its current state is safe, as defined by~\eqref{eq:03:2-5}, then the robot may accelerate with any rate within its physical bounds, cf.~\eqref{eq:03:2-2}.
Finally, if the robot is stopped, it may choose to remain in its current spot and may or may not change its orientation while doing so, cf.~\eqref{eq:03:2-3}.
This is expressed again by arbitrary assignment with subsequent test: this time, the test $?o_r^2=1$, however, restricts the orientation value to either forwards or backwards ($o_r \in \{1,-1\}$).

To stay safe the robot must account for the worst case braking, travel, and acceleration distance, cf.~\eqref{eq:03:2-5}.
This safety margin applies to either the upper or the lower bound of the robot's area, depending on the robot's orientation:
when driving forward (\ie, towards the upper bound), we do not need a safety margin towards the lower bound, and vice versa.
This is expressed by the factors $\tfrac{1-o_r}{2}$ and $\tfrac{1+o_r}{2}$ , which mutually evaluate to zero (\eg, $\tfrac{1-o_r}{2}=0$ when driving forward with $o_r=1$).
The distance between the robot and the obstacle must be large enough to (i) allow the robot to brake to a stand-still, (ii) compensate its current velocity and worst-case acceleration, and (iii) account for the obstacle moving towards the robot with worst-case velocity $V$ while the robot is still not stopped, cf.~\eqref{eq:03:2-6}.
Note, that we have to be more conservative towards the obstacle than towards the bounds, because we want to be able to stop even when the obstacle approaches the robot from behind.

The obstacle, essentially, has similar control options as the robot (with the crucial difference of not having to care about safety): it may either remain in a spot and possibly change its orientation~\eqref{eq:03:3-1}, or choose any velocity up to $V$, cf.~\eqref{eq:03:3-2}.

\subsubsection{Verification}
We verify the acceleration and orientation choices as modeled in~\rref{model:1d} above are safe, using a formal proof calculus for \dL~\cite{DBLP:journals/jar/Platzer08,Platzer10}. 
The robot is safely within its assigned area and at a safe distance to the obstacle, if it is able to brake to a complete stop at all times\footnote{The requirement that the robot has to ensure an option for the obstacle to avoid a collision is ensured trivially, since the obstacle in this model can choose its velocity directly. 
In a more realistic model the obstacle would choose acceleration instead; then the robot had to account for the braking distance of the obstacle, too.}.
The following condition captures this requirement as an invariant \Qte{$r \text{ stoppable } (o,b)$} that we want to hold at all times during the execution of the model:

\begin{align*}
r & \text{ stoppable } (o,b) \equiv~ \norm{x_r - x_o}~\geq~ \frac{v_r^2}{2 B} + \frac{v_o V}{b}
\wedge x_{\underline{b}} + \frac{1-o_r}{2} \frac{v_r^2}{2 B} < x_r < x_{\overline{b}} - \frac{1+o_r}{2} \frac{v_r^2}{2 B}\\
& \qquad \wedge v_r \geq 0 \wedge o_r^2 = 1 \wedge o_o^2 = 1 \wedge 0\leq v_o \leq V\\
\end{align*}

The formula \Qte{$r \text{ stoppable } (o,b)$} states that the distance between the robot to both the obstacle and the bounds is safe, if there is still enough distance for the robot to brake to a complete stop before it reaches either. 
Also, the robot must drive with non-negative velocity, the chosen directions of robot and obstacle must be either forwards ($o_r=1$) or backwards ($o_r=-1$), and the obstacle must use only non-negative velocities up to $V$.

\begin{thm}[Safety of single wheel drive]\label{thm:singlesafety}
If a robot is inside its assigned area and at a safe distance from the obstacle's position $x_o$ initially, then it will not actively collide with the obstacle and stay within its area while it follows the swd control model (\rref{model:1d}), as expressed by the provable \dL formula:
\begin{align}\label{eq:singlesafety_thm}
r \text{ stoppable } (o,b) \rightarrow [\text{\it swd}] \bigl(&(v_r > 0 \rightarrow \norm{p_r-p_o} > 0) \wedge x_{\underline{b}} < x_r < x_{\overline{b}}\bigr)
\end{align}
\end{thm}

We proved Theorem \ref{thm:singlesafety} using \KeYmaera.
With respect to making autonomous systems more scrutable, such a proof may help in a twofold manner: on the one hand, it may increase trust in the implemented robot (given the assumption that the actual implementation and execution can be traced back to the abstract model).
On the other hand, it makes the behavior of the robot more understandable.
In this respect, the most interesting properties of the proven model are the definition of $\textit{safe}$ and the invariant, which allow us to analyze design trade-offs and tell us what is always true about the system regardless of its state.
As an example, let us consider the distance between the robot and the obstacle that is considered safe: $\norm{x_r-x_o} \geq \tfrac{v_r^2}{2 B}+\left(\tfrac{A}{B}+1\right)\left(\tfrac{A}{2} \varepsilon^2+\varepsilon v_r\right) + V\left(\varepsilon+\tfrac{v_r+A \varepsilon}{B}\right)$.
This distance can be interpreted as the minimum distance that the robot's obstacle detection sensors are required to cover (\eg, as done in~\cite{DBLP:conf/iccps/MitschLP12}); it is a function of other robot design parameters (maximum velocity, braking power, worst-case acceleration, sensor/processor/actuator delay) and the parameters expected in the environment (obstacle velocity).
The distance $\norm{x_r-x_o}$ can be optimized \wrt different aspects: for example, to find the most cost-efficient combination of components that still guarantees safety, to specify a safe operation environment given a particular robot configuration, or to determine time bounds for algorithm optimization.

With respect to the manual guidance and collaboration needed in such a proof, we had to apply knowledge in hybrid systems and in-depth understanding of the robot model to find a system invariant, which is the most important manual step in the proof above.
We further used arithmetic interactions, such as the hiding of superfluous terms to reduce arithmetic complexity, transforming and replacing terms (\eg, substitute the absolute function with two cases, one for negative and one for positive values).

\subsubsection{The Proof Structure of Model Variants}
We now want to discuss the proof structure of different model variants:
For example, one can make explicit restrictions on particular variables, such as first letting the robot start in a known direction (instead of an arbitrary direction).
Such assumptions and simplifications, of course, are not without implications on the proof.
While in some aspect a proof may become easier, it may become more laborious or more complex in another.
In this section, we discuss five variants of the single wheel drive model (without obstacle) to demonstrate implications on the proof structure and on the entailed manual guidance needed to complete a proof in \KeYmaera.

The following model variants are identical in terms of the behavior of the robot.
However, assumptions on the starting direction were made in the antecedent of a provable \dL formula, and the starting direction as well as the orientation of the robot were explicitly distinguished by disjunction or non-deterministic choice, or implicitly encoded in the arithmetic, as described below (we denote a changed formula by using primed versions of the original formula reference).

\paragraph{Assumed starting direction, orientation by disjunction.}
In the first variant, the robot is assumed to start in a known direction, specified in the antecedent of \[\tag{\ref{eq:singlesafety_thm}'}\label{eq:variant_1_11} o_r=1\ldots\limply \dibox{\textit{swd}}( x_{\underline{b}} < x_r < x_{\overline{b}}) \enspace .\]
Also, the robot had an explicit choice on turning during stand-still in~\eqref{eq:variant_1_5}. \[\tag{\ref{eq:03:2-3}'}\label{eq:variant_1_5}\textit{ctrl}_r \equiv \ldots \cup (\ptest{v_r=0}; ~\humod{a_r}{0};~ \humod{o_r}{-o_r}) \cup (\ptest{v_r=0};~\humod{a_r}{0})\] 
The orientation of the robot is explicitly distinguished by disjunction in~\eqref{eq:variant_1_6}.
\[\tag{\ref{eq:03:2-5}'}\label{eq:variant_1_6}\textit{safe} \equiv (o_r=-1 \land x_{\underline{b}}+\ldots < x_r) \lor (o_r=1 \land x_r < x_{\overline{b}}-\ldots)\].

\paragraph{Orientation by arithmetic.}
In the second variant, we kept the assumed starting direction of the first variant.
However, the orientation by disjunction in the definition of $\textit{safe}$ was replaced by using $o_r$ as discriminator value encoded in the arithmetic, resulting in~\eqref{eq:03:2-5} of~\rref{model:1d}. 

\paragraph{Arbitrary starting direction by disjunction.}
The third variant relaxes the assumption on the starting direction by introducing a disjunction of possible starting directions in the antecedent of formula~\eqref{eq:variant_1_11} to get~\eqref{eq:variant_3_11}. 
\[\tag{\ref{eq:singlesafety_thm}''}\label{eq:variant_3_11}(o_r=1 \lor o_r=-1)\ldots\limply \dibox{\textit{swd}}(x_{\underline{b}} < x_r < x_{\overline{b}})\]

\paragraph{Arbitrary starting direction by arithmetic.}
The fourth variant replaces the disjunction in the antecedent of~\eqref{eq:variant_3_11} by stating the two orientation options as $o_r^2=1$ to get the antecedent in~\eqref{eq:singlesafety_thm}.

\paragraph{Replace non-deterministic choice with arithmetic.}
Finally, we replace the non-deterministic turning choice of~\eqref{eq:variant_1_5} with $(\ptest{v_r=0};~\humod{a_r}{0};~ \humod{o_r}{*};~ \ptest{o_r^2=1})$ to get~\eqref{eq:03:2-3} of~\rref{model:1d}.
This last variant proves correctness of~\eqref{eq:singlesafety_thm} with $\textit{swd}$ as in~\rref{model:1d}.

\rref{tab:variants} summarizes the proof structures of the five variants.
\begin{table}[tb]
%\centering
\caption{Nodes, branches, and manual proof steps of variants \cite{DBLP:conf/do-form/MitschPP13}}
\label{tab:variants}
\noindent 
\begin{tabularx}{\linewidth}{
  r
  X
  r
  r
  r
  r
  } 
\toprule
  & Variant
  & Nodes
  & Branches
  & Manual steps
  & Avoids
  \tabularnewline \midrule 
(i) & Assumed starting direction, orientation by disjunction & 387 & 34 & 24 & \\
(ii) & Orientation by arithmetic & 331 & 28 & 25 & \eqref{eq:orleft}\\
(iii) & Arbitrary starting direction by disjunction & 650 & 56 & 44 & \\
(iv) & Arbitrary starting direction by arithmetic & 185 & 17 & 22 & \eqref{eq:orleft}\\
(v) & Replace non-deterministic choice & 160 & 14 & 29 & \eqref{eq:choice}\eqref{eq:andright}
  \tabularnewline
  \midrule  
\end{tabularx}\vspace{-2ex}
\begin{tabularx}{\linewidth}{@{}XXX@{}}
\[\tag{$\vee l$}\label{eq:orleft} \frac{\Gamma,\phi\vdash \Delta\quad \Gamma,\psi\vdash\Delta}{\Gamma,\phi \vee \psi \vdash \Delta}\]
&
\[\tag{$[\cup]$}\label{eq:choice} \frac{[a]\phi \wedge [b]\phi}{[a\cup b]\phi}\]
&
\[\tag{$\wedge r$}\label{eq:andright} \frac{\Gamma\vdash\phi,\Delta\quad\Gamma\vdash\psi,\Delta}{\Gamma\vdash\phi\wedge\psi,\Delta}\]
\tabularnewline[-2ex]
\bottomrule
\end{tabularx}
\end{table}
Unsurprisingly---when considering the rules of the \dL proof calculus~\cite{DBLP:journals/logcom/Platzer10} as listed in~\rref{tab:variants}---disjunctions in the antecedent~\eqref{eq:orleft} or in tests of hybrid programs, as well as non-deterministic choices~\eqref{eq:choice} increase the number of proof branches and with it the number of manual proof steps.
Verification experts, who are familiar with the proof calculus and the branching behavior of \KeYmaera, can in some cases express the same model with an alternative encoding using arithmetic expressions.
The number of proof branches can be reduced, if we can replace disjunctions in the antecedent (but also conjunctions in the consequent) or non-deterministic choices in the hybrid program by an equivalent arithmetic encoding.
Conversely, this means that some arithmetic problems can be traded for easier ones with additional proof branches.

\section{Conclusion}
\label{sec:conclusion}

% ------------------------------------------------
% Possible extension stage 2: more details on evaluation, maybe first experience?
% ------------------------------------------------
In this paper, we gave a vision of a verification-driven engineering toolset including hybrid and arithmetic verification tools, and introduced modeling and collaboration tools with the goal of making formal verification of hybrid systems accessible to a broader audience.
The current implementation of \Sphinx\footnote{Available for download at \url{http://www.cs.cmu.edu/~smitsch/sphinx.html}.} includes a textual and a graphical modeling editor for \dL; automated transformation from the graphical models to textual models; simulation of the valuations of variables in hybrid programs over time; integration of \KeYmaera as a hybrid systems verification tool; model and proof comparison; and connection to various collaboration backend systems.

We applied the tool suite to examples based on prior case studies on robotic obstacle avoidance~\cite{DBLP:conf/rss/MitschGP13}.
We made the following two main observations on the effects of collaboration.
\begin{description}
\item[Arithmetic verification] Although less manual effort was put into guiding the automated branching of \KeYmaera, the effort for reducing arithmetic goals to a manageable size for quantifier elimination procedures was reduced by prior collaboration involving geometric relevance filtering~\cite{ppz:grf:2013}.
\item[Domain model decisions] The effect of some modeling decisions---which may come natural to domain experts, such as using disjunction or non-deterministic choice for modeling logical or program alternatives---on the verification effort can be reduced by verification experts that are familiar with the proof calculus by introducing an alternative arithmetic encoding.
\end{description}

As a vision of extending collaboration support, it is planned to integrate Wikis and other online collaboration tools (currently, we use Redmine both as project management repository and for knowledge exchange) for exchanging knowledge on proof tactics. 
Additionally, collaboration with experts outside the own organization can be fostered by linking to Web resources, such as MathOverflow\footnote{\url{mathoverflow.net}}.
In such a platform, requests could be forwarded to those experts whose knowledge matches the verification problem best.

Another interesting research direction are refactoring operations, which systematically construct incremental model variants without the need for re-verification of the entire model~\cite{DBLP:conf/fm/MitschQP14}.
%Refactoring operations are systematic changes applied to a hybrid system model or the properties that we want to prove about them.
In a naive way, after a refactoring was applied we would have to reprove all properties about a model.
But often a refactoring operation changes only fragments of a model while it leaves the remainder of the model untouched, or the refactored model and properties are systematically reducible to previous proofs by side deduction.
A refactoring operation should systematically reduce verification effort by creating new artifacts that are less effort to prove than the complete model.
In a verification-driven engineering process, a refactoring operation creates (i) a refactored model and properties with links to the original model and a description of the applied refactoring;
(ii) a correctness conjecture, together with verification tickets in the project management backend.
This way, all changes applied to models and their properties via refactoring operations can be provably traced back.

The VDE toolset is currently being tested in a collaborative verification setting between Carnegie Mellon University, the University of Cambridge, and the University of Edinburgh.

% *******************************************************************************

\section*{Acknowledgments}

This material is based upon work supported by the National Science
Foundation under NSF CAREER Award CNS-1054246, NSF EXPEDITION
CNS-0926181, and under Grants CNS-1035800 and CNS-0931985, by
DARPA under agreement number FA8750-12-2-0291, and by the US
Department of Transportation's University Transportation Center's TSET
grant, award\# DTRT12GUTC11. Passmore was also supported by the UK's
EPSRC [grants numbers EP/I011005/1 and EP/I010335/1].
This work was also supported by the Austrian Federal Ministry of Transport, Innovation and 
Technology (BMVIT) under grant FFG FIT-IT 829598, FFG BRIDGE 838526, and FFG Basisprogramm 
838181.
The research leading to these results has received funding from the People Programme 
(Marie Curie Actions) of the European Union's Seventh Framework Programme (FP7/2007-2013) 
under REA grant agreement n$^\circ$ PIOF-GA-2012-328378.

\bibliographystyle{abbrv}
\bibliography{proofide}

\end{document}